\documentclass[10pt,aps,pra,twocolumn,
  floatfix %,noshowpacs
  ]{revtex4-2}
\usepackage{float}

\usepackage{amssymb} % {amsfonts} will also be loaded

\usepackage{extarrows} % {amsfonts} already has extendable arrows

\usepackage{dcolumn} % Align table columns of numbers on decimal point

\usepackage{graphicx} % include PDF/EPS/PNG/etc files (for figures, etc)
\usepackage{xcolor}

 \usepackage[normalem]{ulem}

\usepackage{newtxtext,newtxmath}

\newcommand{\yb}[1]{\textsuperscript{#1}Yb}
\begin{document} %%

\title{Quantum correction to the Langevin cross section in resonant-exchange processes}

\author{I. Simbotin}

\author{R. C\^ot\'e}

\affiliation{Department of Physics, University of Connecticut,
 196A Auditorium Rd.,  U-3046 Storrs, CT 06269}

\affiliation{ Department of Physics, University of Massachussetts
  Boston, 100 Morrissey Blvd., Boston, MA 02125-3393}

\date{\today}

\begin{abstract}   %\color{blue}

  Resonant-exchange scattering plays a key role
  in many-body dynamics and transport phenomena (such as spin, charge,
  or excitation diffusion) at low and moderate temperatures.  Recent investigations
  have shown that the locking of phase shifts is central to
  resonant-exchange scattering at low energies.  Moreover, phase-shift locking causes
  an unexpected behavior, namely the resonant-exchange cross section over a broad range of
  energies is largely dictated by $s$-wave scattering, whose influence
  extends high above the $s$-wave Wigner regime.  Here, we generalize
  our previous treatment to higher energies and derive an analytical expression
  for the resonant-exchange cross section which depends on a few 
  parameters and accounts not only for
  the locking of phase shifts, but also for their gradual unlocking with increased energy.  
  We find good agreement between the computed
  (fully quantal) cross sections and those obtained with our expression, which
  we illustrate in great detail for the case of resonant
  charge-transfer in ion-atom collisions.
\end{abstract}

\maketitle

\section{Introduction}

Adapting classical and semi-classical treatments to explain quantal regimes and behaviors 
at ultracold temperatures has a long history. For example, early on, the Wentzel-Kramers-Brillouin 
(WKB) was applied to near-resonant optical excitation of colliding ultracold atoms~\cite{Julienne-1989}, 
paving the way to the prediction and observation of ground-state ultracold 
molecules~\cite{Cote1997-mol,Cote1999-mol,Knize-1998,Pillet-1998}. Quantum suppression and reflection~\cite{Heller-1996,Friedrich-1997,Segev-1998} with respect to semi-classical expectations have been 
used to investigate cold collisions and obtain universal rates \cite{Julienne-Rate-2010,Bohn-2020} for reactions
of ultracold molecules~\cite{Bohn-2022,Bohn-2023b}. A recent review on ultracold molecular collisions \cite{Soley2025} describes
various classical and semi-classical approaches and their level of success. Several reviews on 
ultracold systems cover a range of topics, from collisions to molecules formation and their 
scattering~\cite{Julienne-RMP-2006,Carr-NJP-2009,Dulieu-Review-2011,Julienne-ChemRev-2012,Cote2014-Adv-Chem,Cote-resonance-2017,Bohn-2023a}.

In this article, we are interested in resonant-exchange processes, which involve 
two asymptotically degenerate states and can be understood as the interference of two interaction
paths~\cite{PRL-2018-Simbotin-Cote}.  This well known phenomenon plays an important role in the
behavior of ultracold samples, {\it e.g.}, in spin-exchange in collisions of ultracold atoms and
molecules \cite{yan2013observation, JILA-Progress-Report-2017,JILA-atom-diatom-spin,
ospelkaus2010quantum} or hyperfine-rotation exchange with Rb+KRb \cite{Ni-NatChem-2025}. 
Exchange and interference are actively explored, {\it e.g.}, in experiments on atom-exchange reactions
between NaK Feshbach molecules and K atoms \cite{rui2017controlled} or the role of
interference in chemical reaction \cite{Ni-Science-2024}. 
Excitation resonant-exchange, such as $S$--$P$ excitation exchange for identical
atoms~\cite{Forrey-2000,Bouledroua-2001}, are also central in ultracold Rydberg 
dynamics~\cite{Rydberg-RevModPhys.82.2313} where excitation exchange is involved in F\"oster resonances
\cite{Rydberg-RevModPhys.82.2313,Blockade-PRL}, or quantum random walk
\cite{QRW-Rydberg}.

Resonant-exchange occurs also in systems involving charged particles, such as in  ion-atom
mixtures, where resonant exchange spin-flip dynamics in Yb$^+$+$^{87}$Rb~\cite{Ratschbacher-2013}, 
Yb$^+$+$^6$Li~\cite{Tomza-2018}, and $^{88}$Sr$^+$+Rb~\cite{Sikorsky-2018,Timur-arXiv-2018,Ozeri2025} 
have been studied. In these hybrid systems, because of the extended
range of the ion--atom interaction, the Wigner regime dominated
by $s$-wave scattering is still outside the reach of most
experimental techniques \cite{Ratschbacher-2013,Tomza-2018,Sikorsky-2018,Timur-arXiv-2018,Ozeri2025,vladan09, kohl2010a, denschlag2010,
  kohl2010b,Ratschbacher-2013,Haze-2015, Idziaszek-2011,
  Tomza-2015,Gacesa-Be-2017}, and many partial waves are expected to contribute
 as in charge transfer between an ion and its
neutral parent atom \cite{Cote-Dalgarno-2000,Cote-2000-mobility}.
 However, rapid progress takes place \cite{Cote-review-ion,RMP-ion}, some
 studies involving ultracold molecular ions~\cite{Puri-NatChem-2019}, others ion-atom of the same 
 species like the more recent results on Rb + Rb$^+$~\cite{Katz2025}.

%\clearpage %  jumps to a new page (not just a new column)

%\newpage % break current column;
%   \pagebreak % same as \newpage

%
%\color{violet}%
%

The paper is organized as follows. We first present a general overview of
resonant-exchange collisions in Sec.~\ref{sec:theory}, expressing the quantal 
cross section in terms of its $s$-wave contribution and a classical contribution multiplied 
by a so-called quantal correction function; we describe the classical contribution by the 
Langevin’s capture model and give the simple expression of its cross section for long 
range power-law potentials. In Sec.~\ref{sec:quantal-cross-section}, we give the
expression for the resonant-exchange quantal cross section, apply it to the case of
resonant charge-transfer between an ion and its parent atom, and provide a preliminary discussion  
of both cross section and quantal correction function. We use our scattering phase shift integral 
representation in Sec.~\ref{sec:sum-ell} to truncate the sum over partial waves and find
an approximate expression for the remaining contribution from the phase shifts difference, identifying
the locking and unlocking regimes of phase shifts. Based on those results, we derive an expression
for the quantal correction function in Sec.~\ref{sec:F} based on its general definition, obtain
an explicit approximate equation, and discuss its low and high energy limits. We compare
the results with the exact quantal quantities in Sec.~\ref{sec:results}, paying special attention to the 
locking and unlocking regimes, to finally conclude in Sec.~\ref{sec:conclusion}.

\section{Theoretical overview of resonant-exchange collisions }

\label{sec:theory}

\subsection{General introduction
%    \\ \color{gray}(Generalities)
}

\label{sec:theory-intro}

The behavior of a quantum mechanical system is described by the
Schr\"odinger equation, the physical solutions
obeying suitable boundary conditions, and in the case of scattering
problems, yielding cross sections which can be
analyzed in detail and compared with measurements.  
In this article, we consider the classical counterparts of 
physical quantities derived from a quantal treatment, such as
scattering cross sections, and investigate instead how these classical quantities
are modified by ``quantal corrections.'' 
%
%In this work, we take the point of view that it
%can be advantageous to consider the classical counterparts of the
%physical quantities derived from a quantal treatment, such as
%scattering cross sections, and to regard them as classical quantities
%modified by ``quantal corrections.''  
%\sout{Such correction factors (or
%functions) can simply be employed as convenient tools when
%analyzing experimental data, or can be given a more rigorous
%underpinning based on a fully quantal treatment. }
%{\color{red}\sout{While the latter is our goal, 
%our framework and fitting procedures also provide a pragmatic approach.}}

\begin{figure}[t]
  \centerline{
    \includegraphics[width=0.99\linewidth]{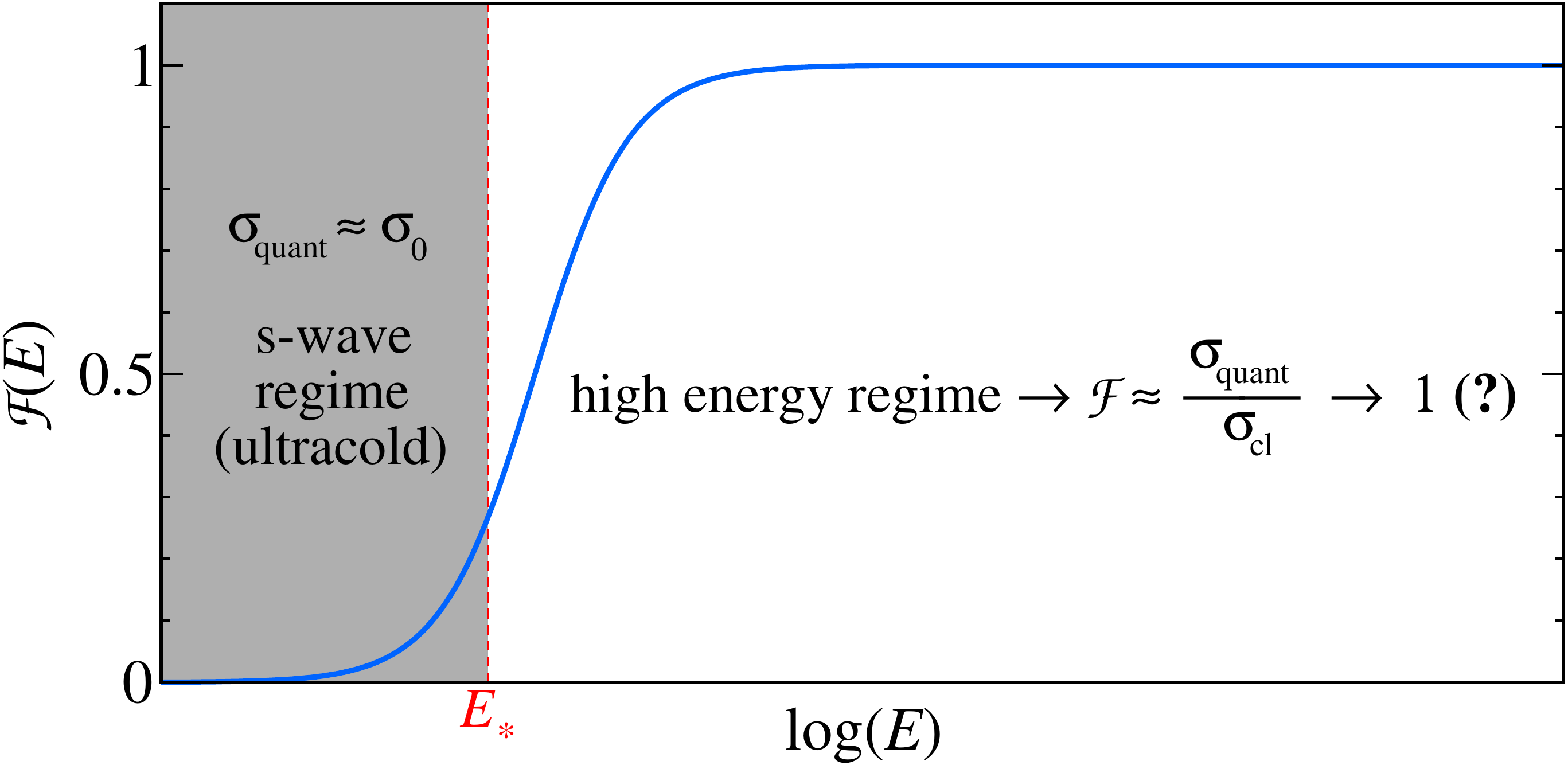}
%    \framebox[0.9\linewidth][c]{\rule[-7ex]{0pt}{15ex}
%      ...... Sketch for $F(E)$ vs. $\log(E)$..........}
  }
  \caption{
    \label{fig:F-sketch}
    Schematic energy dependence of the quantal correction function,
    assuming that the classical treatment is adequate at high energy,
    which would yield $\mathcal F\approx1$; however, as indicated by
    the question mark, the validity of the classical approximation
    must be assesed by comparing it with the quantal approach.  The
    vertical dashed line marks the boundary (denoted $E_*$) of the
    s-wave regime of ultralow energies (emphasized by shading).  }
\end{figure}

We first consider the cross
section for a generic scattering process.  As we stated above, one can
regard the exact (fully quantal) cross section ($\sigma_{\rm quant}$)
as its classical counterpart ($\sigma_{\rm cl}$) modified by a quantal
correction function (denoted $\cal F$) which is defined according to
\begin{equation}
  \sigma_{\rm quant}(E) = \sigma_0 (E) + {\cal F}(E) \sigma_{\rm cl} (E),
  \label{eq:F-generic}
\end{equation}
where $E$ is the energy of the relative motion of the collision
partners and $\sigma_0$ is the quantal $s$-wave cross section.  The
latter is singled out because it is the dominant contribution at
vanishingly low energies, where the classical picture becomes
inadequate and a quantal treatment is necessary.  Conversely, for
energies above the $s$-wave regime, $\sigma_0(E)$ will become
negligible and thus Eq.~\eqref{eq:F-generic} reduces to a simpler
definition of the quantal correction function:
$\mathcal{F}\approx\frac{\sigma_\text{quant}}{\sigma_\text{cl}}$.
Therefore, if the classical treatment yields a cross section that is
in agreement with the quantal result, $\mathcal F$ is expected to
approach unity at higher energies.  Consequently,
$\mathcal{F}(E)$ can have a rather simple energy dependence, as
sketched in Fig.~\ref{fig:F-sketch}, provided that the classical
picture is suitable.  However, we shall see that quantal effects can
be significant not only at low energies but also at ``high'' energies,
resulting in a nontrivial energy dependence which may depart
substantially from the behavior depicted in Fig.~\ref{fig:F-sketch}.
Finally, we need to clarify a matter of language; namely, as indicated
in Fig.~\ref{fig:F-sketch}, the terminology ``high energy'' will be
used relative to the ultralow energy scale $E_*$ (defined in Section~\ref{sec:power-law} 
based on the long-range power-law tail of a general interaction
potential).

\subsection{Langevin's capture model}
\label{sec:Langevin}

\begin{figure}[t]
\centerline{  \includegraphics[width=0.9\linewidth]{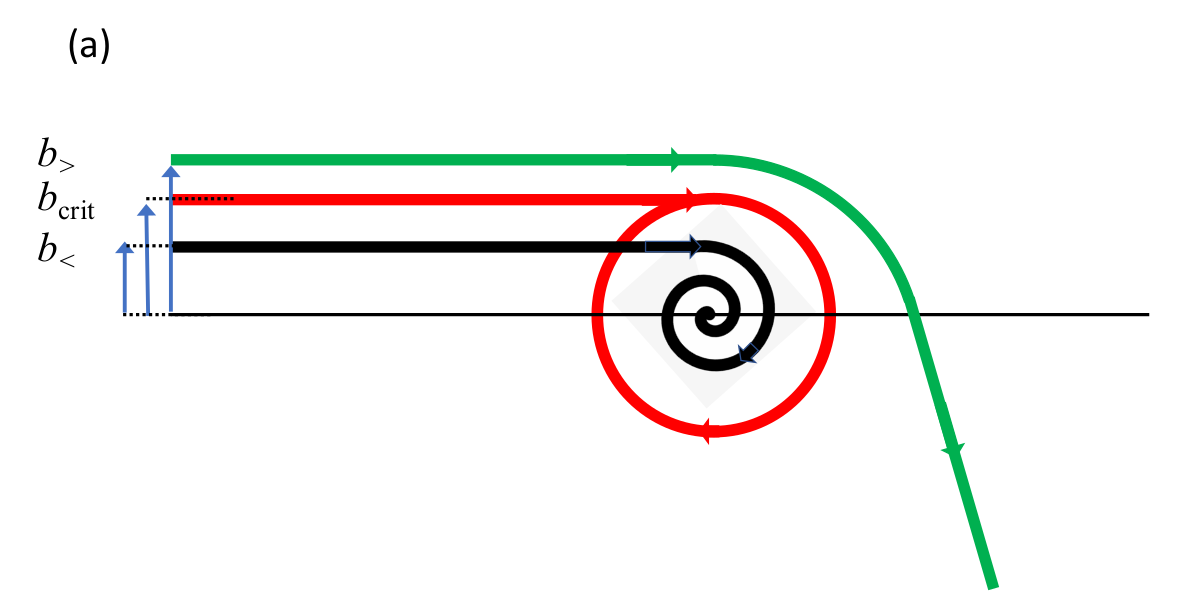}}
\centerline{  \includegraphics[width=0.9\linewidth]{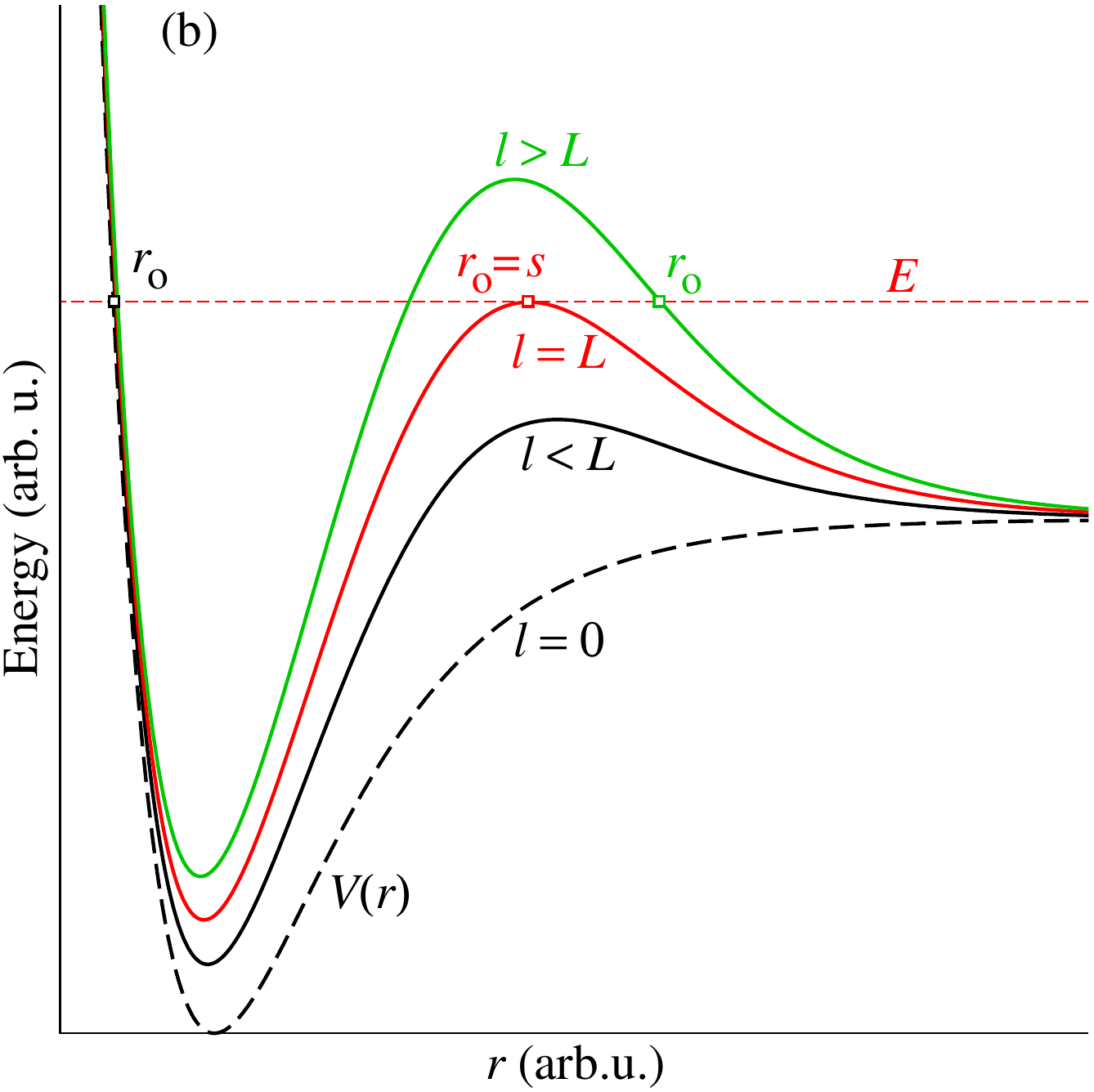}}
\caption{\label{fig:capture-sketch}
   Schematic depiction of Langevin's capture model. The upper plot (a)
   shows classical trajectories for three impact parameters; for small
   values of the impact parameter, the projectile follows a trajectory
   (black curve labeled $b_<$) spiraling into the short-range region,
   where it will be captured by the target and the ``reaction'' will
   take place.  For large values, the trajectory (green curve labeled
   $b_>$) is deflected away and the projectile will not reach the
   immediate vicinity of the target; thus, the reaction does not occur.
   The critical case corresponds to an orbiting trajectory (red curve
   labeled $b=b_\text{crit}$).  The lower plot (b) shows a typical
   interaction potential energy curve, $V(r)$ depicted by the dashed
   line, along with the effective potentials for the three cases in
   the upper panel.  The barrier is due to the centrifugal term,
   $\frac{\ell(\ell+1)\hbar^2}{2\mu r^2}$, which is included in the
   effective potential.  For a given collision energy, $E$, the
   critical value $\ell=L(E)$ of the angular momentum corresponds to
   $b=b_\text{crit}$, while $\ell<L$ and $\ell>L$ correspond to $b_<$
   and $b_>$, respectively.  The outermost turning point $r_\text{o}$
   is shown for each potential curve; in the critical case
   $r_\text{o}$ coincides with the location, $s$, of the top of the
   centrifugal barrier.}
\end{figure}

In this work, we focus on  resonant-exchange scattering,
for which the classical picture yields a Langevin-type
cross section (denoted $\sigma_{L}$ and defined below).  As
illustrated in Fig.~\ref{fig:capture-sketch}(a), the Langevin
approximation \cite{Langevin1905,mott-massey,Cote-Dalgarno-2000,Cote-review-ion,RMP-ion,Newton_2002,Taylor1972,Joachain_1975} relies on the key
assumption that any classical trajectory with an impact parameter $b$
smaller than a critical value ($b_{\rm crit}$) will spiral in towards
short distances where the projectile is captured by the target and the
transition process (reaction, or state change) takes place with unit
probability, while for $b>b_{\rm crit}$ there is no capture and hence
no reaction (because the short-range region remains out of reach).
The critical value of the impact parameter, $b=b_\text{crit}$,
corresponds to an orbiting trajectory.

In the classical picture, the impact parameter reads $b= J/p$, with $J$
and $p$ the angular and linear momenta of the relative motion, while
in the quantal treatment $J=\hbar\sqrt{\ell(\ell+1)}$, where $\ell$ is
the angular momentum quantum number which labels the scattering
partial waves, and $p=\hbar k=\sqrt{2\mu E}$ with $k$ the wave number
and $\mu$ the reduced mass of the projectile-target system.  In order
to bridge the classical and quantal pictures, we make use of the
high-$\ell$ approximation,
$J\approx\hbar(\ell+\frac{1}{2})\approx\hbar\ell$, and we thus write
$b=\ell/k$, which will be used below.

The effective potential, $V_\text{eff}(r)=V(r)+\frac{J^2}{2\mu r^2}$,
involved in the capture process mentioned above is depicted
schematically in Fig.~\ref{fig:capture-sketch}(b).  We remark that the
centrifugal term,
$\frac{J^2}{2\mu{}r^2}=\frac{\hbar^2\ell(\ell+1)}{2\mu{}r^2}$, creates
a barrier of increasing height and width as $\ell$ increases.  For low
values of $\ell$, the barrier remains below $E$ and thus the
short-range region is accessible; this is where couplings to other
potential curves allow for a change in the internal state, e.g., an
exchange process or ``reaction''.  The classical capture picture
relies on the assumption that the reaction takes place with unit
probability if the short-range region is reached.  As $\ell$
increases, the barrier will grow to reach a critical height equal to
$E$, and thus the outermost classical turning point $r_\text{o}$ will
coincide with the position (denoted $s$) of the top of the barrier;
the critical value of $\ell$ is denoted $L(E)$ which is energy
dependent and is allowed to vary continuously.  As mentioned above,
the critical case corresponds to an orbitting trajectory in the
classical picture, and we thus have $L\approx{}kb_\text{crit}$.  For
$\ell>L$, the barrier will be higher than $E$ and will prevent the
system from reaching the inner region, thus precluding the ``reaction"
from taking place; also, as $\ell$ increases, the outermost classical
turning point $r_\text{o}$ will shift farther into the outer region,
which is a clear indication that the short-range region is out of
reach.  However, it is well known that in the quantal approach
tunneling through the barrier is possible for $\ell>L$, which will
give rise to resonant features (shape resonance) in the energy
dependence of the cross section.

According to the brief discussion above, we now write the expression
of the Langevin cross section in the classical capture picture
\cite{Langevin1905,mott-massey,Cote-Dalgarno-2000,Cote-review-ion},
\begin{equation}
  \sigma_L(E) = \pi b_{\rm crit}^2 =  \frac{\pi}{k^2} L(L+1)
  \simeq  \frac{\pi}{k^2} L^2(E),
    \label{eq:Langevin}
\end{equation}
where we employed $b^2_{\rm crit}=L(L+1)/k^2\approx L^2/k^2$.  We shall
refer to Eq.~(\ref{eq:Langevin}) as the \emph{generalized} Langevin
cross section, since we have not yet used any specific assumptions for
the interaction potentials.  In the next section we adopt the typical
asymptotic behavior of interatomic potentials in order to find the
energy dependence for $L(E)$ and hence $\sigma_L(E)$.  The latter 
is substituted as the classical cross section $\sigma_\text{cl}(E)$ in
Eq.~(\ref{eq:F-generic}) which we rewrite as
\begin{equation}
  \sigma_\text{exc}(E) = \sigma_0 (E) + {\cal F}(E) \sigma_L(E),
  \label{eq:F-exact-intro}
\end{equation}
where $\sigma_\text{exc}$ is the quantal cross section for
resonant-exchange scattering (to be defined explicitly in
Sec.~\ref{sec:theory-sigma-exc}).

\subsection{  Long range power-law potentials }
\label{sec:power-law}

As is well known, the interaction potentials for atomic and molecular
systems behave asymptotically ($r\to\infty$) as a sum of inverse-power
terms,  $V(r)\approx\sum_n V_n(r)$ \cite{Dalgarno58b,Cote-review-ion,PRL-2018-Simbotin-Cote}, with
\begin{equation}\label{eq:V=CnRn}
V_n(r) = \pm C_n r^{-n}.
\end{equation}
%where $n>2$.
For the sake of convenience, we shall take into account only the
dominant term, i.e., the lowest power $n$ (e.g., $n=4$ for polarization potential in ion--atom
systems or $n=6$ for the van der Waals potential between neutral ground state atoms).  Such an  interaction potential (with an
asymptotic tail behaving as a pure inverse power)  makes it possible
to obtain a simple expression for the Langevin cross section.
We note however that quantal results presented in this work were computed using the realistic full potential curves.

For scattering at very low energies, capture processes can only take
place if the interaction is attractive in the asymptotic region
($r\to\infty$), i.e., $V(r)\sim V_n(r)=-C_nr^{-n}$ with $C_n>0$.
The power-law exponent $n$ and the coefficient $C_n$ dictate the
strength of the long-range interaction, and they can be used together
with $\hbar$ and the reduced mass $\mu$ to define a length scale,
\begin{equation}\label{eq:R-star}
   R_*\equiv\left(2\mu\hbar^{-2}C_n\right)^{\frac1{n-2}},
\end{equation}
which yields a momentum scale, $\hbar k_*=\hbar/R_*$,
and hence an energy scale,
\begin{equation}\label{eq:E-star}
E_* = \frac{\hbar^2k_*^2}{2\mu}
    = \frac{\hbar^2}{2\mu R_*^2}
    = \frac{C_n}{R_*^n}.
\end{equation}
For $n=6$, the quantities $R_*$ and $E_*$ are referred to as the van
der Waals length and energy scale, respectively; by extension, this
terminology is sometimes also used for systems whose interactions
follow other ($n\ne6$) power laws.  Typical values for $R_*$ and $E_*$
are given below in Sec.~\ref{sec:theory-example} to show that they are
the relevant scales for the s-wave regime of ultralow energies, as
indicated in Fig.~\ref{fig:F-sketch}.  Moreover, it is well known that
the domain of vanishingly low energies is characterized by cross
sections behaving according to Wigner's threshold laws~\cite{Wigner1948}
and is refered to as the Wigner regime (which we consider synonymous
with the ultracold regime, defined conventionally as $0<E\lessapprox
E_*$).
%
%thus, $E_*$ is the appropriate energy scale for the Wigner (threshold) regime.

The range of energies explored in our work will be restricted to low
and moderate $E$, to ensure that the effective potential for partial
waves $\ell\leq L(E)$ has an inner well delimited by a centrifugal
barrier, as sketched in Fig.~\ref{fig:capture-sketch}(b).  Moreover,
the energies are sufficiently low such that the long-range tail of the
interaction plays a dominant role in the scattering process; in
particular, we would like to ensure that the centrifugal barrier for
$\ell\leq L(E)$ remains outside the short-range region, in order for
the interaction potential to be well approximated by its asymptotic
expression~\eqref{eq:V=CnRn}.  Anticipating our discussion of the
quantal cross section and the partial-wave expansion, we mention that
$L(E)$ will be justifiably used as a cutoff for the partial-wave
series, as will be explained in Sec.~\ref{sec:truncation}.  At much
higher energies, which are outside the scope of our investigation,
$L(E)$ will continue to grow and the centrifugal term will eventually
become dominant and will lift the potential well completely.  More
specifically, in the high-$\ell$ limit, the effective potential will
become a monotonic function of $r$, as the inner well and the barrier
merge together and disappear; consequently, $L(E)$ cannot be defined
any longer at high $E$, and Langevin's approach will be rendered
inapplicable.
%(or it will simply yield a value of zero for the cross section).  
Therefore, to ensure that Langevin's capture model remains
applicable and $L(E)$ is well defined, the scope of our work is
restricted to low-energy scattering on interaction potentials which
are asymptotically attractive.  The extent of the energy range
explored in this study will be made specific for a concrete example in
Sec.~\ref{sec:theory-example}.

A simple expression for the cutoff $L(E)$ can be obtained by making
use of the asymptotic form~(\ref{eq:V=CnRn}) of the interaction
potential and imposing the condition for the critical height of the
barrier, $E=-\frac{C_n}{s_L^n}+\hbar^2\frac{L(L+1)}{2\mu s_L^2}$,
where
$s_\ell\equiv\Big(\frac{nC_n\mu}{\ell(\ell+1)\hbar^2}\Big)^\frac1{n-2}$
is the position of the top of the centrifugal barrier along the radial
coordinate for a given $\ell$.  The result can be written more conveniently if we employ
a redefined cutoff, $\Lambda \equiv L (L+1)$, which reads
\begin{equation}\label{eq:LLn}
%  L(L+1) =
  \Lambda(n;E) 
 = g_n \left(\frac E{E_*}\right)^{\!\frac{n-2}n}\!\!
 = \frac{2\mu}{\hbar^2} g_n \left(C_n\right)^{\frac{2}n} E^{\frac{n-2}n},
\end{equation}
with the numerical factor
\begin{equation}\label{eq:gn}
g_n=\frac{n}{n-2}\left(\frac{n-2}2\right)^{\frac2{n}}
\end{equation}
obeying $1<g_n\leq 2$ for $n>2$.  We now substitute 
Eq.~(\ref{eq:LLn}) in Eq.~(\ref{eq:Langevin}) to obtain the
\emph{generalized} Langevin cross section,
\begin{equation}\label{eq:sigma-L-n}
  \sigma_L (E) \equiv  \frac{\pi}{k^2} \Lambda (E)
   = \pi R_*^2 g_n \left( \frac{E_*}{E}\right)^{\frac2{n}}
  = \pi g_n\left(\frac{C_n}E\right)^\frac2{n}.
\end{equation}
In Table~\ref{tab:sigma_L}, we give the explicit expression for $R_*$, $E_*$, and $\sigma_L (E)$ for the most common power-law $n=3$, 4, and 6.
\begin{table}
\caption{``Van der Waals" length $R_*$ and energy $E_*$, and Langevin cross section $\sigma_L$ for various $n$.}
\label{tab:sigma_L}
\begin{tabular}{r|cccccc} \hline\hline %&&&&&&\\
 $n$ && 3 & & 4 & & 6  \\ \hline
  %&&&&&\\ 
$R_*$ && $\displaystyle \frac{2\mu}{\hbar^2}C_3$
           && $\displaystyle \left(\frac{2\mu}{\hbar^2}C_4\right)^{1/2}  $
           && $\displaystyle \left(\frac{2\mu}{\hbar^2}C_6\right)^{1/4} $  \\
$E_*$ && $\displaystyle \frac{\hbar^6}{8\mu^3 C_3^2}$ 
           && $\displaystyle \frac{\hbar^4}{4\mu^2 C_4}$  
           && $\displaystyle \frac{\hbar^3}{2\sqrt{2}\mu^{3/2} \sqrt{C_6}}$   \\
$ \sigma_L $ &  &$\displaystyle 3 \pi  \left(\frac{ C_3}{2E}\right)^{2/3}$ 
   && $\displaystyle 2\pi \left(\frac{C_4}{E}\right)^{1/2}$
   && $\displaystyle \frac{3\pi}{2}\left(\frac{2 C_6}{E}\right)^{1/3}$ \\ 
  %&&&&&&\\ 
  \hline\hline
\end{tabular}
\end{table}

% We emphasize that
The power-law energy dependence of the Langevin cross section is
rather simple and is therefore expected to only have limited
validity, as we will show in this work.
% it can differ significantly---even drastically---from the quantal cross section.

\section{The cross section}
\label{sec:quantal-cross-section}

\subsection{The quantal cross section for resonant-exchange scattering}

\label{sec:theory-sigma-exc}

The resonant-exchange collision (or reaction) for an atomic species
$X$ with internal states $\alpha$ and $\alpha'$ can be written in a
generic fashion as
\begin{equation}
  X^\alpha + X^{\alpha'} \longrightarrow X^{\alpha'} + X^\alpha.
\label{eq:exchange-process-general}
\end{equation}
Typical examples include spin-flip collisions with spin states
$\alpha=|\uparrow\,\rangle$ and $\alpha'=|\downarrow\,\rangle$, or
excitation-exchange between $\alpha =S$ and $\alpha' = P$ atomic
states, or charge-transfer collisions between a positively charged ion
and its neutral parent atom with their electric charges $\alpha=+1$
and $\alpha'=0$ given in units of the elementary charge $q=|e|$.

A qualitative picture for resonant-exchange scattering can be grasped
in terms of amplitude interference of two reaction paths $a$ and $b$
with interaction potentials $V_a(r)$ and $V_b(r)$ which
become degenerate asymptotically ($r\to\infty$).  More specifically,
the cross section for such a process can be written as a partial-wave
sum involving the scattering phase shifts $\eta_\ell^{a/b}$ for the
potential curves $V_{a/b}$
\cite{Cote-Dalgarno-2000,Cote-review-ion,mott-massey},
\begin{equation}
   \sigma_{\rm exc} (E)= \frac{\pi}{k^2} \sum_{\ell =0}^\infty (2\ell +1) 
   \sin^2\left(\Delta\eta_\ell\right),
\label{eq:csec-def}
\end{equation}
with $\Delta\eta_\ell$ the phase-shift difference
\begin{equation}\label{eq:Delta-eta}
\Delta\eta_\ell \equiv \eta_\ell^b - \eta_\ell^a.
\end{equation}
The generic labels $a$ and $b$ denoting the reaction paths (channels)
correspond to molecular electronic states (e.g., gerade and ungerade,
or singlet and triplet) which are asymptotically degenerate, as we
shall see in the next section where we discuss a concrete example.

%\clearpage%\newpage

\subsection{Example: resonant charge-transfer}

\label{sec:theory-example}

\begin{figure}[b]
\includegraphics[width=\linewidth]{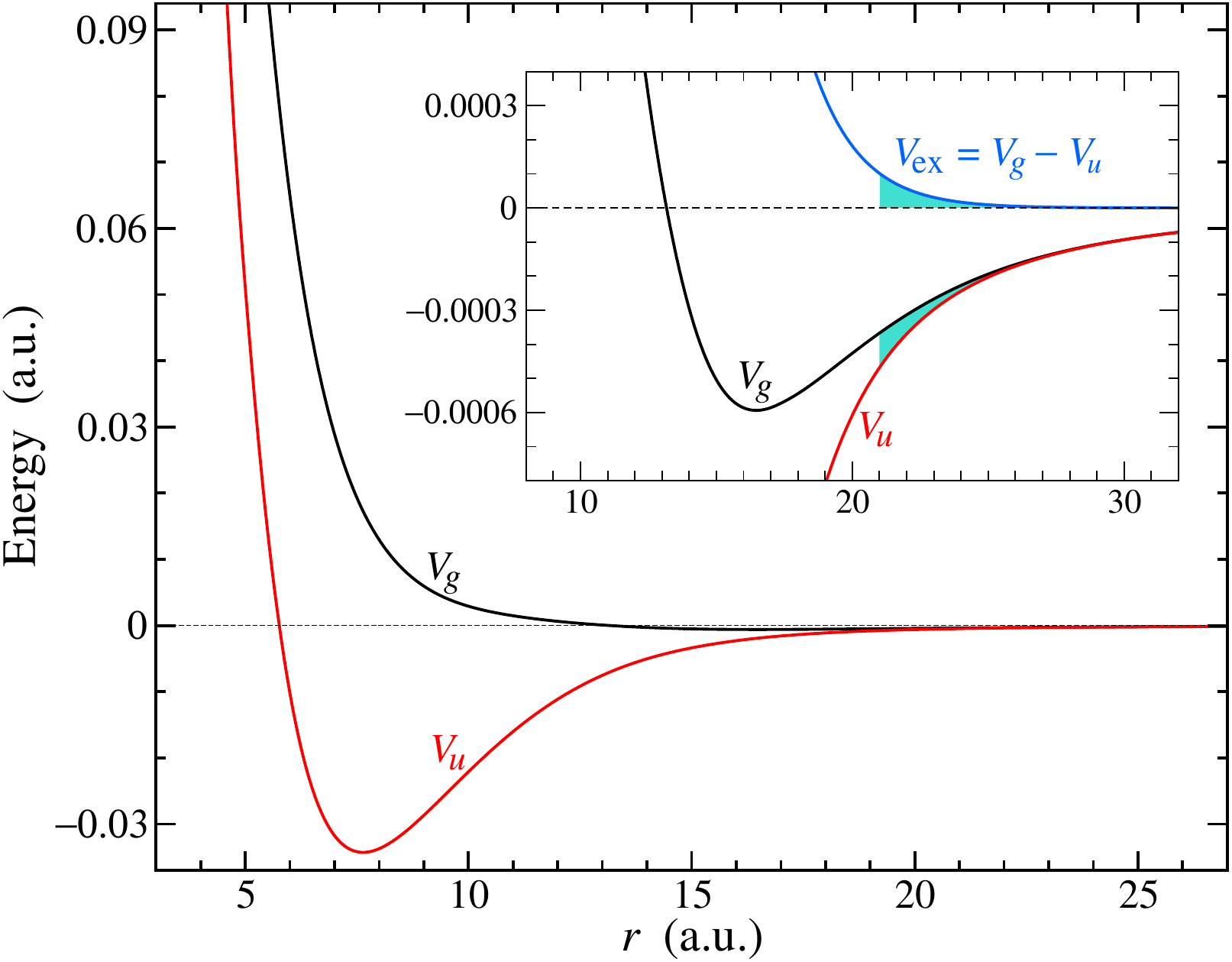}
\caption{\label{fig:Yb-potentials} Potential curves for the lowest
  electronic states $^2\Sigma_{u/g}^+$ of Yb\,--Yb$^+$.  The lower
  potential curve ($V_u$) has a deep minimum at
  $R_u^\text{min}\approx7.6$~a.u., while the upper curve ($V_g$) has a
  very shallow well located at $R_g^\text{min}\approx16.5$~a.u., as
  shown in the inset.  Shadding is used to emphasize the rapid
  vanishing of the exchange term of the interaction,
  $V_\text{ex}=V_g-V_u$, which becomes negligible for $r>R=30$~a.u.,
  and thus the two potential curves become nearly identical for
  $r>R$\@.  We remark that the value of the potential energy at the
  boundary ($r=R$) separating the short- and long-range regions is a
  significant fraction of the depth of the shallow potential,
  $V_g^\text{depth}\approx6\times10^{-4}$~a.u., which is many orders
  of magnitude higher than the energy scale of the $s$-wave regime;
  indeed, $V_{u,g}(R)\sim10^{-4}$~a.u., while $E_*\sim10^{-13}$~a.u.
}
\end{figure}

We choose the charge-transfer between an ion and its parent neutral
atom as a case study for our investigation of resonant-exchange
scattering;  our illustrative system is ytterbium, chosen for its many isotopes
leading to a range of behaviors \cite{Yb-PRA-2009}. Thus, the
generic collision process in Eq.~\eqref{eq:exchange-process-general} reads
%($X=\text{Yb}$)
\begin{equation}
%  X + X^+ \longrightarrow X^+ + X.
    \text{Yb} + \text{Yb}^+ \longrightarrow \text{Yb}^+ + \text{Yb}.
\label{eq:exchange-process-yb}
\end{equation}
The  potential curves \cite{Yb-PRA-2009} for the two lowest electronic states of Yb$_2^+$
are shown in Fig.~\ref{fig:Yb-potentials}.  Note that the channel
labels ($a$, $b$) introduced earlier correspond to \emph{ungerade}
($a=u$) and \emph{gerade} ($b=g$) symmetries of the
$^2\Sigma_{u/g}^+$  states involved in the scattering process.  As is apparent
in Fig.~\ref{fig:Yb-potentials}, the two potential curves become
degenerate at long-range ($r\gtrapprox30$~a.u.) and take on the
asymptotic behavior $V_u(r)=V_g(r)\approx-C_4/r^4$ with
$C_4=\alpha_d/2$, where $\alpha_d=145$~a.u.\@ is the dipole static
polarizability of the neutral Yb atom.  The long-range tail of the
interaction plays a very important role in low-energy scattering and
will be further discussed in Sec.~\ref{sec:truncation}.

%\newpage

The energy range explored in this work extends from $E\rightarrow 0$ to high
above the Wigner regime ($E\ggg E_*$) but is restricted to
$E<V_\text{depth}\approx6\times10^{-4}$~a.u.,
% ~a.u.\@ (equivalently, $V_\text{depth}/k_B\sim190$~K)
with $V_\text{depth}$ the depth of the shallow potential curve $V_g$.
More specifically, to ensure the validity of certain approximations
that we shall employ, our assumption will be $E\ll V_{\rm depth}$,
which is adequately fulfilled for energies
$E\lessapprox10^{-4}$~a.u.\@ (temperatures $T=E/k_B$ lower than 30~K).  As we
already mentioned, these energies are sufficiently low such that the
long-range tail of the interaction plays a dominant role in the
scattering process.  We emphasize that the short-range energy scale is
many orders of magnitude higher than the energy scale of the Wigner
regime, i.e., $V_\text{depth}\ggg E_*$; indeed, using the atomic mass
of $^{172}$Yb, we obtain
$E_*=1.4\times10^{-13}$~$\text{a.u.}\approx3.8\times10^{-12}$~$\text{eV}\approx44$~nK,
which is exceedingly low.  Such a value for $E_*$ is typical for
ion-atom collisions and stems from the long-range nature of their
interactions, for which the length scale $R_*$ is very large in
general; e.g., for Yb--Yb$^+$, Eq.~(\ref{eq:R-star}) yields
$R_*=4.8\times10^3$~a.u.~(Bohr radii).

\subsection{Preliminary discussion of results}

 \label{sec:prelim}

In order to clarify the motivation behind our work, we give here a
brief summary of our results, while a detailed discussion will be
presented in Sec.\kern0.1em\ref{sec:results}\kern-2pt\@.  We first
rewrite Eqs.~(\ref{eq:LLn}) and (\ref{eq:sigma-L-n}) for the cutoff
$\Lambda=L(L+1)$ and the  Langevin cross section in the
particular case of ion-atom collisions, i.e., $n=4$.  Making use of
the coefficient $g_n$ defined in Eq.~(\ref{eq:gn}), which attains its
maximal value ($g_4=2$) for $n=4$, the cutoff $\Lambda$ in
Eq.~(\ref{eq:LLn}) reads
\begin{equation}\label{eq:LL4}\color{black}
  \Lambda(E) = 2kR_* = 2 \sqrt{\frac E{E_*}}
  = \frac{4\mu}{\hbar^2} \sqrt{C_4 E},
\end{equation}
while the generalized expression of $\sigma_L$ in
Eq.~(\ref{eq:sigma-L-n}) yields the well known Langevin cross section
for ion-atom scattering,
\begin{equation}\label{eq:sigma-L-4}\color{black}
 \sigma_L (E) = \frac{2\pi R_*}k = 2\pi\sqrt{\frac{C_4} E}.
\end{equation}

\begin{figure}[t]
\includegraphics[width=\linewidth]{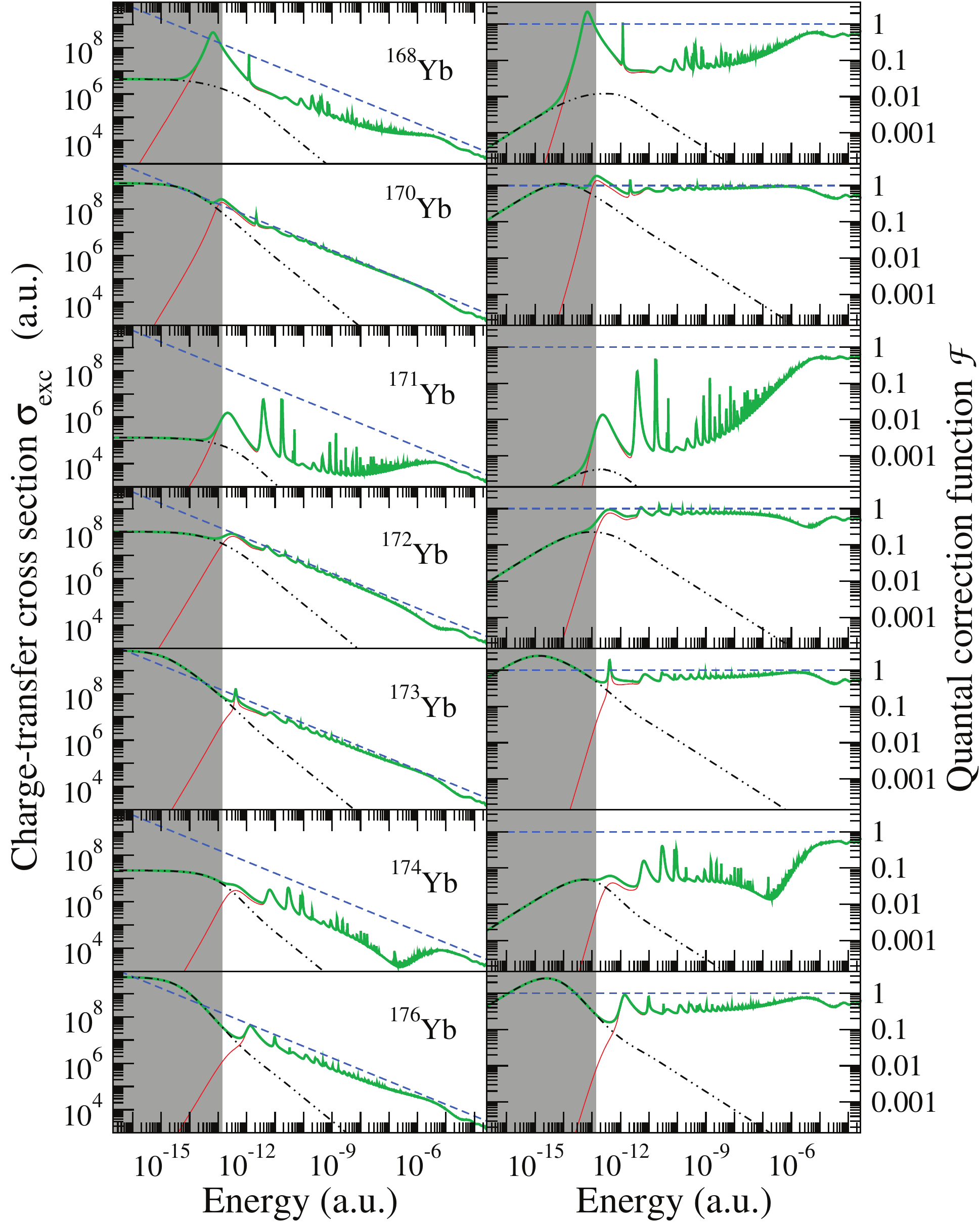}
\caption{Left side: resonant charge-transfer cross sections for
  ion-atom collisions for all isotopes of Yb.  Quantal cross section
  $\sigma_{\rm exc}$ (green, solid line), $s$-wave ($\ell=0$)
  contribution $\sigma_0$ (black, dash-dot line), $\ell\geq 1$
  contribution (red, solid line), Langevin cross section $\sigma_L$
  (blue, dashed line).  Right side: the corresponding quantal
  correction functions ${\cal F}$ (red line), ratios $\sigma_{\rm
    exc}/\sigma_L$ (green line) and $\sigma_0/\sigma_L$ (dash-dot
  line), and the Langevin approximation represented by the horizontal
  dashed line.  The shaded areas represent the Wigner regime of
  ultralow energies ($E<E_*$) where s-wave scattering is dominant.
}
\label{fig:Yb-sigmas-F-ratio}
\end{figure}

\begin{figure}[t]
  \includegraphics[width=\linewidth]{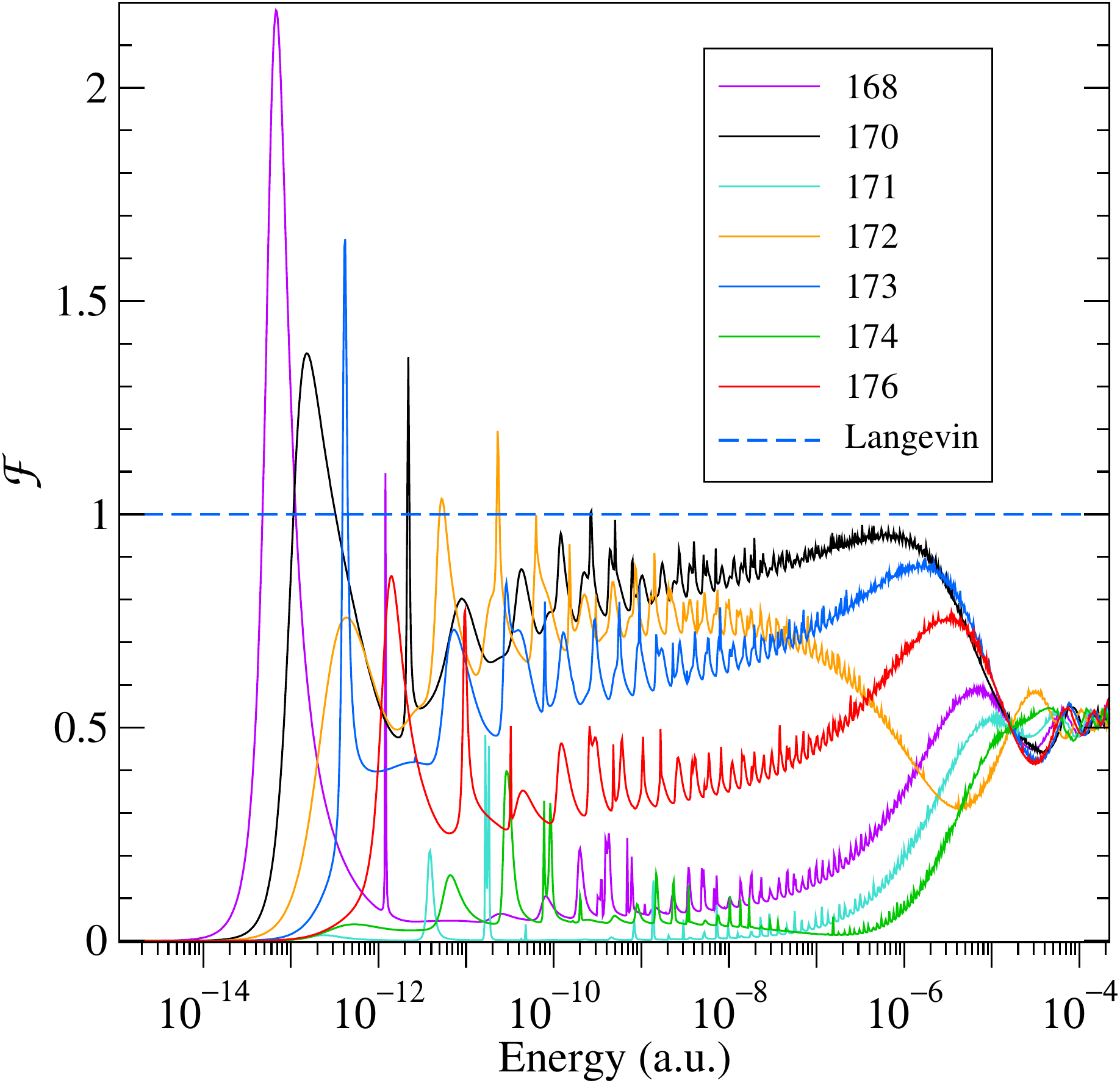}
\caption{
  The quantal correction function, $\mathcal{F}(E)$, for all Yb
  isotopes; the Langevin approximation is depicted by the horizontal
  dashed line at unity (as in Fig.~\ref{fig:Yb-sigmas-F-ratio}).  Note
  the linear scale of the vertical axis, while the horizontal (energy)
  axis uses a logarithmic scale (as in
  Fig.~\ref{fig:F-sketch}).  
  Note that the computed results shown in this plot represent the
  exact quantal correction function, which differs substantially from
%%  and it is instructive to compare and  contrast them with
  its expected behavior based on the classical Langevin model (see
  Fig.~\ref{fig:F-sketch}).
  %% \textcolor{magenta}{ shown here as the thick \uwave{\bfseries
  %%     magenta \bfseries curve}} \color{teal} or maybe drop it?...
%
  %% \color{gray}  
  %% Also note that the size (height) of the sharp features (shape
  %% resonances) is about the same for all isotopes, even though in
  %% Fig.~\ref{fig:Yb-sigmas-F-ratio} they seem to stick out much higher
  %% above the background; this confusion is due to the different scales
  %% used for the vertical axis (log- and linear-scale) in the two
  %% figures.  In Fig.~\ref{fig:Yb-sigmas-F-ratio}, the low background in
  %% supression cases allows the sharp features (shape res) to be more
  %% visible because the log-scale plot is roughly equivalent to plotting
  %% ratios (spike height relative to background) on a linear scale; a
  %% lower background yields a higher ratio.
}
\label{fig:F-exact-semilog}
\end{figure}

Figure~\ref{fig:Yb-sigmas-F-ratio} displays our computed 
(fully quantal) cross sections for  resonant
charge-transfer collisions of an ion with its parent atom for all
isotopes of Yb.  Also shown are the individual contribution for
$s$-wave ($\ell=0$) and the cumulative contribution for all other
($\ell\geq 1$) partial waves, as well as the Langevin cross section
$\sigma_L(E)$.  The corresponding quantal correction function
$\mathcal{F}(E)$ is shown in the panels on the right side of
Fig.~\ref{fig:Yb-sigmas-F-ratio}, and also depicted in greater detail
as a semilog plot in Fig.~\ref{fig:F-exact-semilog}.  For the sake of
clarity, we recast Eq.~(\ref{eq:F-exact-intro}) as an explicit
definition of the quantal correction function,
\[
\mathcal{F} \equiv\frac{\sigma_\text{exc}-\sigma_0}{\sigma_L}
=\frac{1}{\sigma_L}  \sum_{\ell=1}^\infty\sigma_\ell,
% \label{eq:F=def-exact} % in case this eq will be numbered/labeled
\]
where $\sigma_\ell=\frac\pi{k^2}(2\ell+1)\sin^2(\Delta\eta_\ell)$ are
the partial-wave terms of the quantal cross section.
%$\sigma_\text{exc}$ in Eq.~(\ref{eq:csec-def}).

It is readily apparent in Fig.~\ref{fig:Yb-sigmas-F-ratio} that the
Langevin cross section diverges at vanishingly low energies, where a
fully quantal treatment is required and $s$-wave scattering is
dominant (yielding a finite cross section,
$\sigma_\text{exc}\approx\sigma_0$).  At higher energies, one would
expect $\sigma_L(E)$ to reach agreement with the exact cross section
$\sigma_\text{exc}(E)$.  However, our results show that strong quantal
effects persist to energies which are many orders of magnitude above
the $s$-wave regime; indeed, despite the fair agreement for isotopes
$^{170}$Yb, $^{172}$Yb and $^{173}$Yb, the quantal results for all
other isotopes exhibit considerable discrepancy with the Langevin
approximation over the entire energy domain.  We must conclude that
$\sigma_L$ is in general not a good approximation to the 
exact cross section $\sigma_\text{exc}$.
%Thus, the first general
%conclusion is that   there is a rather poor agreement between
%$\sigma_L$ and the exact cross section $\sigma_\text{exc}$.

Figure~\ref{fig:Yb-sigmas-F-ratio} also shows that the $s$-wave term
($\sigma_0$) becomes negligible  above the Wigner regime, as the
contribution of higher ($\ell\geq1$) partial waves begins to dominate.
Thus, $\sigma_0$ can be neglected for $E\gtrsim E_*$, yielding
$\mathcal{F}\approx\frac{\sigma_\text{exc}}{\sigma_L}$, which greatly
simplifies the Langevin picture when expressed in terms of the quantal
correction function.  Indeed, as depicted  earlier in
%
%\marginpar{\footnotesize\bfseries\color{teal}Fig.1}%
%
Fig.~\ref{fig:F-sketch}, the validity of the Langevin approximation
would correspond to $\mathcal{F}$ attaining a value of unity above the
$s$-wave regime, i.e.,
\[
\mathcal{F}\approx\frac{\sigma_\text{exc}}{\sigma_L}
\quad \xLongrightarrow[\text{approx.}]{\text{Langevin}}
\quad \mathcal{F} = 1.
\]

%\newpage

The value yielded by the Langevin approximation ($\mathcal F=1$) is
represented by the horizontal dashed line in the panels on the right
side of Fig.~\ref{fig:Yb-sigmas-F-ratio} (and also in
Fig.~\ref{fig:F-exact-semilog}) and facilitates the comparison of the
Langevin capture model with the quantal approach. 
The quantal correction function $\mathcal F$ is plotted on a logarithmic 
scale in Fig.~\ref{fig:Yb-sigmas-F-ratio} to emphasize the $s$-wave and
$\ell \geq 1$ contributions, while Fig.~\ref{fig:F-exact-semilog} depicts all
of them in the same plot on a linear scale to accentuate their range of variability and
the impact of shape resonances.
 Moreover, the
Langevin cross section seems to be an upper bound for the quantal
cross section, which is readily apparent
% in our detailed comparison
in Fig.~\ref{fig:Yb-sigmas-F-ratio}.  The upper bound interpretation,
i.e., the inequality $\sigma_\text{exc}<\sigma_L$ (or its equivalent
 version, $\mathcal{F}<1$, as indicated in Fig.~\ref{fig:F-exact-semilog})
will be justified in our theoretical analysis in the subsequent
sections.  However, there are some exceptions at very
low energy where $\sigma_\text{exc}>\sigma_L$ due to shape resonances
in low partial waves, including resonance enhancement in $s$-wave
inside the Wigner regime for isotopes $^{173}\text{Yb}$ and
$^{176}\text{Yb}$.

Finally, we remark that high above the Wigner regime the shape
resonances become ubiquitous and merge into a smooth background, which
can exhibit a striking variety of behavior (as is  clearly visible
in Fig.~\ref{fig:F-exact-semilog}).  We will focus our attention on the
energy dependent background, while ignoring the shape resonances
(whose impact near threshold is well understood; e.g., see~\cite{Simbotin_2014,Simbotin_2015,Shu_2017,Cote-resonance-2017}).  Consequently, our main goal will be to
explain the smooth features of the energy dependence of the quantal
results, which requires a detailed analysis of the partial-wave
expansion~(\ref{eq:csec-def}) and the scattering phase shifts.
We note that resonant charge-transfer has a long history of approximations 
over the years ({\it e.g.}, see \cite{Dalgarno58b,Sakabe1992,Simon-Manso2023}).

%%%%%%%%%%%%

\section{Partial wave expansion}

\label{sec:sum-ell}

In order to understand the behavior of the quantal correction function
(and hence the cross section), we first analyze the individual
phase shifts and their difference,
$\Delta\eta_\ell=\eta_\ell^b-\eta_\ell^a$.  
%\sout{Next, the partial-wave sum
%will be converted into an integral which can be evaluated in closed
%form, provided that we make use of an approximation for the $\ell$
%dependence of $\Delta\eta_\ell$.  This will yield an approximate
%expression for $\mathcal{F}$ which accounts remarkably well for the
%overall behavior of the quantal cross section across a wide range of
%energies.}

%\newpage

\subsection{ Integral representation of the scattering phase shifts}

\label{sec:eta=int-rho}

We  now analyze the phase shifts using an integral representation which we 
developed~\cite{pra-2018-int,pra-2019-barrier} based on a
phase-amplitude approach~\cite{Milne_1930,Boisseau-Julienne-2000}.  
Assuming an energy-normalized wave function of the form 
$\psi_\ell(r) = \sqrt{\rho_\ell (r)/k}\sin[\theta_\ell (r) - \theta_\ell (0)]$, 
the phase shift $\eta_\ell$ for a given
scattering potential, $V(r)$, can be expressed as
\begin{equation}
  \eta_\ell = \ell \frac{\pi}{2} + \int_0^\infty dr
  \left[ \frac{k}{\rho_\ell (r)} - k \right],
   \label{eq:eta_ell-original}
\end{equation}
where the envelope function $\rho_\ell(r)$ is the solution of the
third order linear diﬀerential equation
$\rho_\ell ''' - 4U_\ell  \rho_\ell ' - U_\ell ' \rho_\ell =0$
together with the asymptotic boundary condition 
$\rho_\ell(r)\rightarrow 1$ when
$r\rightarrow \infty$, where 
$ U_\ell  = \frac{2\mu}{\hbar^2} (V-E) + \frac{\ell (\ell +1)}{r^2} $.
We emphasize that
the integral representation above yields the true value of the
phaseshift $\eta_\ell$ (not modulo $\pi$); a full account of our phase-amplitude
formalism can be found in
Refs.~\cite{pra-2018-int,pra-2019-barrier} 
%
%, as we discuss in
%Appendix~\ref{appendix:wkb}.  Suffice to say that
%Eq.~(\ref{eq:eta_ell-original}) is an exact expression, and thus
%$\frac{\hbar k}{\rho(r)}$ can be regarded as the quantal (exact)
%version of the local classical momentum and is well defined everywhere
%(including the classically forbidden regions).

\begin{figure}[b]
   \includegraphics[width=\linewidth]
                   {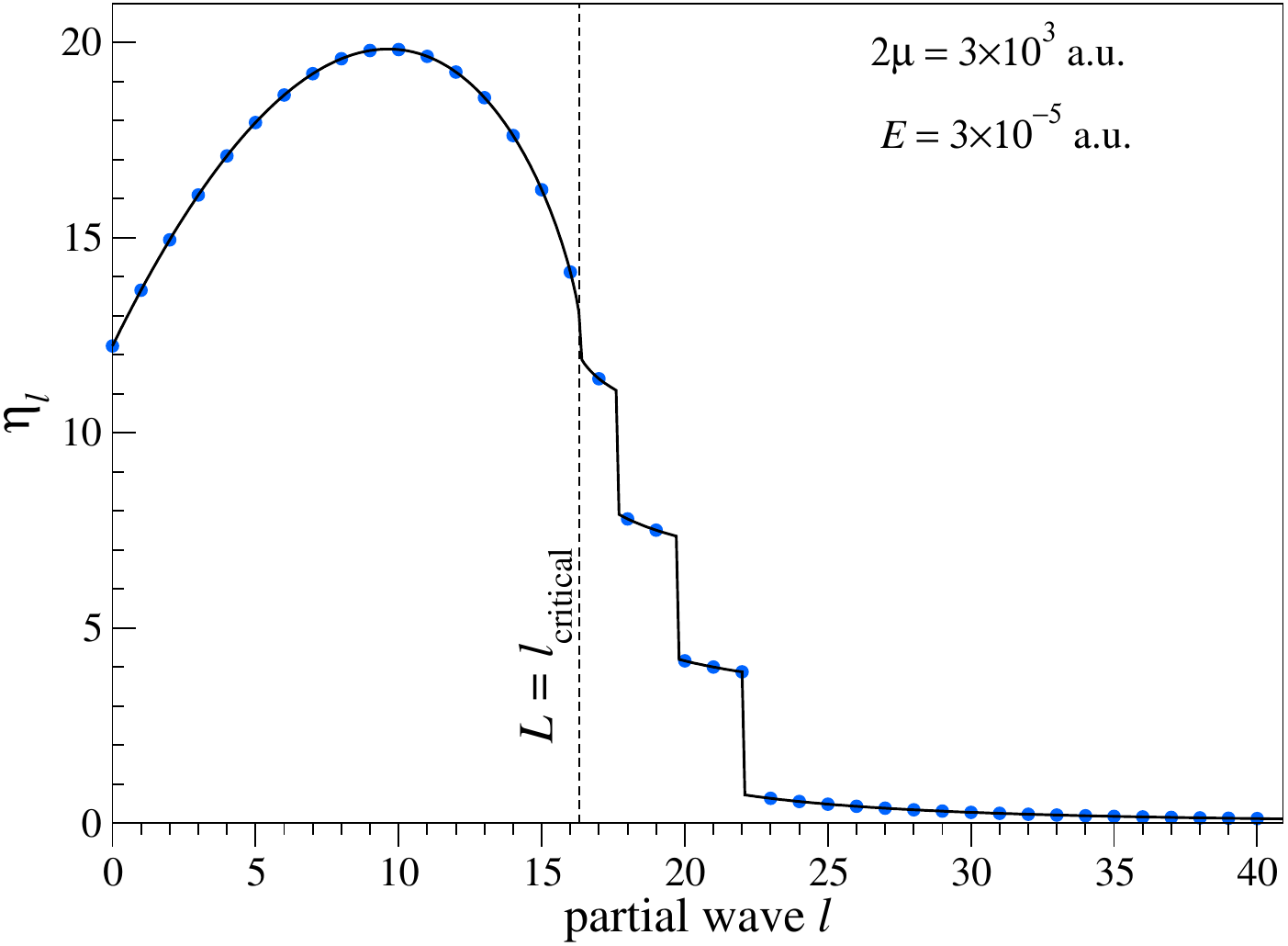}
 \caption{\label{fig:eta_ell}
  Partial wave dependence of the scattering
  phase shift $\eta_\ell$  for the potential
  $V_g$ of the state $^2\Sigma_g^+$ of Yb$_2^+$ for a fixed collision
  energy, $E=3\times 10^{-5}$ a.u.  For illustrative purposes and convenience, we used an
  artificial (much lower) value of the reduced mass,
  $\mu=1.5\times10^3$ a.u.  The full (black) curve was obtained by
  allowing $\ell$ to be continuous, while $\eta_\ell$ for integer
  values of $\ell$ are indicated by the blue dots.  The vertical
  dashed line marks the critical value of the partial wave,
  $\ell_\text{critical}=L$, for which the height of the centrifugal
  barrier equals the scattering energy $E$.
}
\end{figure}

    %\newpage

We first discuss key features of the $\ell$ dependence of
$\eta_\ell$ at a fixed energy, as illustrated in
Fig.~\ref{fig:eta_ell} for a fixed energy $E$ and a fictitious reduced mass 
$\mu=1.5\times 10^3$ a.u. (for illustrative purposes only); the $E$ dependence will be explored later.  
As is readily apparent in Fig.~\ref{fig:eta_ell}, the
critical partial wave ($L$) marks the boundary between
two different types of behavior.  For $\ell<L(E)$ [see Fig.~\ref{fig:capture-sketch}(b)], 
the height of the centrifugal barrier is lower than the collision energy,
$E$, the wave function extends to the inner region of the potential,
and the phase shift has a smooth behavior.
For partial waves higher than the crtical value, $L(E)$, the 
barrier is higher than $E$ and shape resonances may appear; indeed,
for $\ell>L$, sharp steps are clearly visible in
Fig.~\ref{fig:eta_ell}, which correspond to shape resonances held by
the potential well at short range 
%\cite{mott-massey,Bernstein_1960,pra-2019-barrier}.  
\cite{mott-massey,pra-2019-barrier}. Note also that
$\eta_\ell$ vanishes in the limit $\ell\to\infty$, as expected.  We
remark that, as $\ell$ increases, the fixed energy $E$ can be regarded
as \emph{decreasing} relative to the increasing height of the
centrifugal barrier (which is responsible for shape resonances).
Hence, as $\ell$ increases,  $\eta_\ell$
\emph{decreases} by $\pi$ for each resonance.

\subsection{ Truncation of the partial-wave sum}
  \label{sec:truncation}

For $\ell > L(E)$, the top of the centrifugal barrier located at
$r=s$ [see Fig.~\ref{fig:capture-sketch}(b)]) is higher than $E$ and a point $s'\approx s$ 
provides a natural separation of the
inner ($r<s'$) and outer ($r>s'$) regions.  Accordingly, as shown in
Ref.~\cite{pra-2019-barrier}, the contributions of the two regions
can be separated by splitting the integration domain at $r=s'$.  Thus,
the phase shift in Eq.~(\ref{eq:eta_ell-original}) reads
\begin{equation}
  \eta_\ell(E) =  \pi N_\ell(E) + \eta_\ell^\text{out}(E)
  \qquad \text{for }\; \ell>L(E),
      \label{eq:eta-split-N-tail}
\end{equation}
with the two terms defined as follows:
\begin{gather}
  \pi N_\ell  \equiv  \int_0^{s'} \!\! dr \frac{k}{\rho_\ell (r)},\label{eq:N-pi=int}
  \\[2.ex]
  \label{eq:eta-tail}
\eta_\ell^\text{out}  \equiv  \ell\frac{\pi}{2} -ks' + \int_{s'}^\infty \!\!dr
\left[ \frac{k}{\rho_\ell (r)} -k \right].
\end{gather}
We note that the location of $s'$ in the vicinity of $s$ is chosen such that 
Eq.(\ref{eq:N-pi=int}) is satisfied, {\it i.e.} $N_\ell$ is an integer equals to the number of 
nodes of the wave function $\psi_\ell(r)$ in the inner region.
The second term, $\eta_\ell^\text{out}(E)$, is the outer-region
contribution which has a smooth (albeit nontrivial) dependence on $\ell$
and $E$ and accounts for the smooth behavior of $\eta_\ell$
in between the sharp steps in Fig.~\ref{fig:eta_ell}.  The first term,
$\pi N_\ell(E)$, is the phase accumulated inside the inner region and
amounts to an integer multiple of $\pi$, which corresponds to the
number of bound and quasi-bound ro-vibrational levels below $E$.  Thus,
provided that $\ell>L(E)$, $N_\ell(E)$ behaves in a stepwise fashion
as a function of $\ell$ (and also as a function of $E$), which is due
to shape resonances held by the inner potential well at short-range.
As mentioned above, for a fixed energy $E$ below the top of the
barrier, $N_\ell$ exhibits a \emph{downward} step for each resonance
as $\ell$ increases, as illustrated in Fig.~\ref{fig:eta_ell}.
  Conversely, for a fixed value of $\ell$, the
energy dependence of $N_\ell(E)$ would show an abrupt increase by
unity when $E$ sweeps through each shape resonance.
If the energy would continue to increase, it would approach (and then
surpass) the top of the barrier, and thus the resonances would begin to
broaden and eventually disappear into a smooth background.

We remark that in the case of elastic
scattering~\cite{mott-massey,Cote-Dalgarno-2000,Cote-resonance-2017} the contribution of high
partial waves ($\ell>L$) is substantial, while for resonant-exchange
scattering they have a negligible contribution;  indeed, the
partial-wave sum~\eqref{eq:csec-def} can be safely truncated at
$\ell\approx L$, as we show next.

We now turn our attention to the phase shift \emph{difference},
$\Delta\eta_\ell=\eta^b_\ell-\eta^a_\ell$, where the labels $a,b$
correspond to the potential curves $V_{a,b}$ of the electronic states
involved in the resonant-exchange scattering process.  We take
advantage of the fact that the two electronic states are
asymptotically degenerate and their potential curves at large
separation can be written as
\[
V_{a,b}(r) \rightarrow V_\text{asympt}(r) \mp\tfrac1{2} V_{\rm ex}(r).
\]
The exchange term, $V_{\rm ex}(r)\sim r^\alpha\exp(-\beta r)$
\cite{Bardsley_1975,Cote-Dalgarno-2000,Cote-review-ion}, vanishes
exponentially outside the short-range region, and thus the remaining
interaction at large distances is $V_\text{asympt}$.  As is well
known, the latter can be written as an asymptotic series of inverse
power terms;  in our case (resonant charge exchange for atom-ion
scattering) the long-range interaction reads
\[
V_\text{asympt}(r) = -\frac{C_4}{r^4} - \frac{C_6}{r^6}- \frac{C_8}{r^8} - \cdots
\]
The leading term, $V_\text{asympt}(r)\approx-\frac{C_4}{r^4}$,
vanishes slowly and plays an important role even at distances of
thousands of Bohr radii, while the exchange term only plays a role at
short-range, as shown in Fig.~\ref{fig:Yb-potentials}.  Hence, the two
potential curves have identical long-range tails,
\begin{equation}
  \label{eq:Va=Vb}
V_a(r) = V_b(r) = V_\text{asympt}(r)
\quad\text{for}\quad r>R,%=30\text{ a.u.},
\end{equation}
with $R\approx 30$~a.u.~chosen as the boundary separating the short- and
long-range regions for the Yb+Yb$^+$ system (see Fig.~\ref{fig:Yb-potentials}).

Equation~(\ref{eq:Va=Vb}) can now be used to justify the truncation of
the partial-wave sum at $\ell=L$\@.  First, we recall the definition
of $L(E)$ as the critical $\ell$ for which the height of the
centrifugal barrier reaches $E$\@.  More specifically, for each
potential curve $V_{a,b}$ we have $L_{a,b}(E)$, which may be different
($L_a\neq L_b$) in general.  However, we shall keep the collision
energy low enough such that the partial waves $\ell\leq L_{a,b}(E)$
have centrifugal barriers located in the long-range region ($r>R$)
where $V_a(r)=V_b(r)$, thus ensuring
\[
L_a(E)= L_b(E),
\]
which will simply be denoted $L(E)$.  The condition above is typically
fulfilled for energies much lower than the short-range energy scale of
either potential curve, $E\ll V_{a,b}^\text{depth}$.

We now focus on high partial waves ($\ell>L$) and employ
Eq.~(\ref{eq:eta-split-N-tail}) for both potential curves,
\[
\eta^{a,b}_\ell = \pi N^{a,b}_\ell\ +\eta_\ell^\text{out}.
\]
We recall that $\eta_\ell^\text{out}$ stems from the outer region
alone, where the two potential curves are identical and hence the
envelopes are also identical: $\rho^a_\ell(r)=\rho^b_\ell(r)$.
Consequently, we have
$\eta_\ell^{\text{out},a}=\eta_\ell^{\text{out},b}$ and thus the phase
shift difference reduces trivially to an integer multiple of $\pi$,
\begin{equation}
  \Delta\eta_\ell  = \eta^b_\ell-\eta^a_\ell
  = \pi\Delta N_\ell  \quad \text{ for }\; \ell>L ,
   \label{eq:Deta-eta=Npi}
\end{equation}
where $\Delta N_\ell(E)=N_\ell^a(E)-N_\ell^b(E)$ is an integer valued
function of $E$ (except for very narrow shape resonances, which we
omit).  We thus have $\sin^2(\Delta\eta_\ell)=\sin^2(\pi\Delta
N_\ell)=0$ for $\ell>L$, which justifies the truncation of the
partial-wave sum at $\ell\approx L$,
\begin{equation}
   \sum_{\ell =0}^\infty (2\ell +1) 
   \sin^2\Delta\eta_\ell \approx  \sum_{\ell =0}^L (2\ell +1) 
   \sin^2 \Delta\eta_\ell.
\label{eq:csec-def-approx}
\end{equation}
Next, we will analyze the behavior of the phase shifts for the
remaining partial waves ($\ell<L$).

\subsection{ Discussion of phase shifts for $\ell<L$}

\label{sec:eta-ell-A}

To tease out the $\ell$-dependence in $\eta_\ell$ 
from Eq.(\ref{eq:eta_ell-original}), 
we combine the integral representations for
$\ell=0$ and $\ell\neq0$ to write
\begin{equation} 
   \eta_\ell - \eta_0 =  \ell \frac{\pi}{2} - \!\int_0^\infty \!\!dr
    \frac{k\hat \rho_\ell (r)}{\rho_\ell (r)\rho_0 (r)},
   \label{eq:eta_ell-0ref=int}
\end{equation}
where $\hat\rho_\ell\equiv\rho_\ell-\rho_0$. 
By employing the two-envelope approach developed in
our previous work (see Eqs.(33) and (34) in \cite{pra-2018-int}), 
using $\ell=0$ as the reference, with 
$U_{\rm ref}=U_{\ell=0}  = \frac{2\mu}{\hbar^2} (V-E)$ and 
$\rho_{\rm ref}=\rho_0$, and the centrifugal term 
$\hat{U}(r)=\frac{\ell(\ell+1)}{r^2}$ as the ``perturbation'', $\hat \rho_\ell$ 
satisfies \cite{pra-2018-int}
\begin{equation}
   \hat \rho_\ell '''  - 4U_\ell \hat \rho_\ell '  -2U_\ell ' \hat \rho_\ell 
   = 4 \frac{\ell (\ell +1)}{r^2} \left( \rho_0 ' - \frac{\rho_0}{r}  \right).
   \label{eq:DE-delta_rho_ell}
\end{equation}
Since $\rho_0$ and $\rho_0 '$ do not depend on $\ell$, $\hat \rho_\ell$
is proportional to $\ell (\ell +1)$, albeit an additional $\ell$ dependence 
due to the centrifugal term in $U_\ell  = \frac{2\mu}{\hbar^2} (V-E) + \frac{\ell (\ell +1)}{r^2}$. We thus write
\begin{equation}
   \hat\rho_\ell(r) \equiv \ell(\ell+1) \bar\rho_\ell(r) ,
\label{eq:rho-bar}
\end{equation}
where $\bar\rho_\ell(r)$ contains the remaining $\ell$-dependence due to $U_\ell$.

Substituting into Eq.(\ref{eq:eta_ell-0ref=int}) allows to write the individual phase
shifts for channels $i$ as
\begin{equation} 
  \eta^i_\ell(k) = \eta^i_0 (k) + \ell\frac{\pi}{2}
  -\ell(\ell +1) A_\ell^i (k),
   \label{eq:eta_ell-A_ell}
\end{equation}
where $A_\ell^i$ is defined %in Eq.~(\ref{eq:A_ell}) 
in terms of envelope functions $\rho^i_\ell(r)$ and $\bar\rho^i_\ell(r)$,
\begin{equation}
   A^i_\ell (k) =  \int_0^\infty dr 
   \frac{ k\bar\rho^i_\ell(r)}{\rho^i_\ell(r)\rho^i_0(r)} \;. \label{eq:Delta-eta_ell^i}
 \end{equation}
The label $i=a,b$ refers to the potential curves
$V_i$ of the two electronic states involved in the scattering process.
We note that,  according to the Levinson theorem \cite{mott-massey,Harald-book,Newton_2002,Taylor1972,Joachain_1975},
\begin{equation}
  \eta_0^i (k)=N_0^i \pi + \delta_0^i(k)\;,
  \label{eq:delta_0}
\end{equation}
where the integer $N_0^i$ is the number of 
bound levels supported by the potential $V^i$ as $E\rightarrow 0$
and $\delta_0^i$ the residual scattering phase shift. 

Making use of Eq.~(\ref{eq:eta_ell-A_ell}), the phase shift difference
reads
\begin{equation}
  \label{eq:Delta-eta_ell}
  \Delta\eta_\ell(k) \equiv \eta_\ell^b-\eta_\ell^a
= \Delta\eta_0(k)-\ell(\ell+1)\Delta A_\ell(k),
\end{equation}
where $\Delta \eta_0 \equiv \eta_0^b - \eta_0^a$ and $\Delta A_\ell \equiv A^b_\ell - A^a_\ell$.
A similar expression was derived using a semi-classical approach \cite{PRL-2018-Simbotin-Cote}.

As in the previous section, we take advantage of Eq.~(\ref{eq:Va=Vb})
 to simplify the integral representations for $\Delta\eta_\ell$
and for the newly defined $\Delta A_\ell$.  Namely, we rely on the
fact that the envelope functions for scattering problems are specified
by the asymptotic boundary condition $\rho_\ell^i\rightarrow 1$ and
thus they are initialized at $r=\infty$ and propagated inward (see
Ref.~\cite{pra-2018-int} for details).  According to
Eq.~(\ref{eq:Va=Vb}), we have $V_a=V_b=V_\text{asympt}$ throughout the
entire long-range region ($R<r<\infty$), which yields
$\rho_\ell^a=\rho_\ell^b=\rho_\ell^\text{asympt}$ (similarly for $\bar\rho^i_\ell$)
and consequently the
long-range contributions to the individual phase shifts $\eta^i_\ell$
are equal.  Indeed, we simply split the integration domain at $r=R$ in
Eq.~(\ref{eq:eta_ell-original}),
\[
%\begin{equation}
    \eta^i_\ell(k) = \ell \frac{\pi}{2}
    + \int_0^R dr  \left[ \frac{k}{\rho_\ell^i (r)} - k \right] 
    +         \int_R^\infty dr
  \left[ \frac{k}{\rho_\ell^\text{asympt} (r)} - k \right],
  \label{eq:eta_ell-split-R}
%\end{equation}
\]
and we identify the last term as the long-range contribution, which is
identical for the individual phase shifts $\eta_\ell^a$ and
$\eta_\ell^b$.  Thus, the phase shift difference can be written as an
integral over the short-range region alone,
\begin{equation}
\Delta\eta_\ell(k) = \eta^b_\ell - \eta^a_\ell
=\int_0^R dr
\left[ \frac{k}{\rho_\ell^b (r)} - \frac{k}{\rho_\ell^a (r)}  \right].
   \label{eq:Delta-eta_ell-R}
\end{equation}
Similarly, $\Delta A_\ell$ in Eq.~(\ref{eq:Delta-eta_ell}) also
benefits from long-range cancellation; indeed, making use of
Eq.~(\ref{eq:Delta-eta_ell^i}) for $A^i_\ell$, we have
\begin{align} 
   \Delta A_\ell(k)
     & =  A^b_\ell(k) - A^a_\ell(k)    \nonumber
 \\
     &= \int_0^R dr 
  \left[ \frac{k \bar\rho_\ell^a (r)}{\rho_\ell^a (r)\rho_0^a (r)}
        -\frac{k\bar\rho_\ell^b (r)}{\rho_\ell^b (r)\rho_0^b (r)}  \right] .
\label{eq:DA_ell-exact}
\end{align}

\begin{figure}[t]
  \includegraphics[width=0.77\linewidth]
   {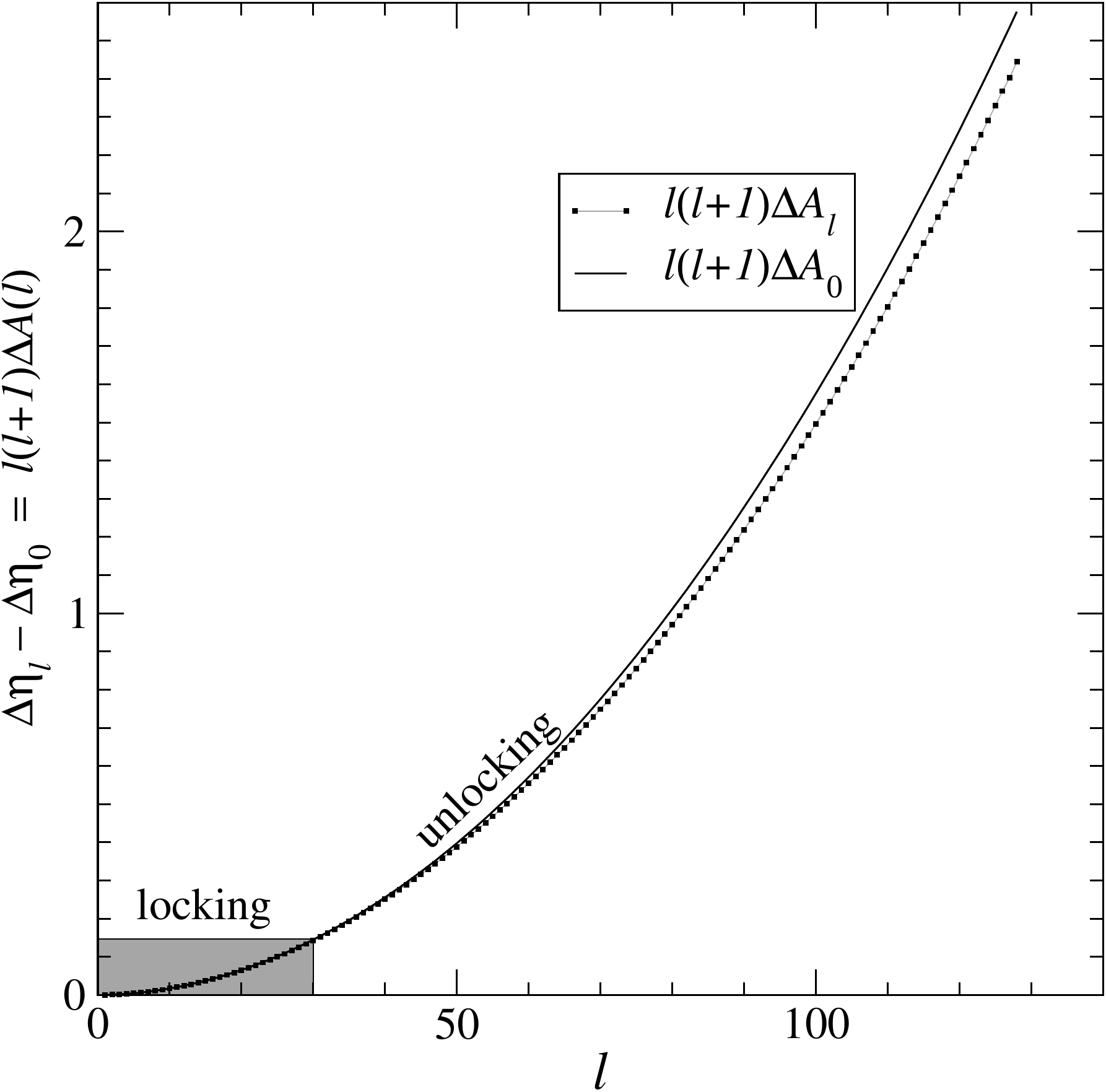}
 % linear plot-vs-ell
   \caption{\label{fig:Delta_eta_ell} The $\ell$-dependence of $\Delta\eta_\ell$ 
     for $\ell<L(E)$ at $E=10^{-5}$~a.u. of $^{170}$Yb, showing the
     nearly quadratic growth with $\ell$. $\Delta\eta_\ell$ varies slowly 
     (less than $\pi$) due to the small value
     $\Delta A_\ell\approx1.6\times10^{-4}$. The small dots stand for the
     quantal (exact) results ($\ell=0,1,2,$ etc.) and the solid 
     curve corresponds to the approximate
     expression~(\ref{eq:Delta-eta_0}).
  The regimes for locking (shaded area) and unlocking of phase shifts are also shown 
  (see Sec.~\ref{sec:eta-lock} and Sec.~\ref{sec:lock}).
}
\end{figure}

In Fig.~\ref{fig:Delta_eta_ell}, we display the contribution $\ell (\ell+1) \Delta A_\ell (k)$ 
for a specific $k$ corresponding to $E=10^{-5}$ a.u. for $^{170}$Yb. In particular, we find
that term varies slowly for $\ell<L$, increasing by less than $\pi$.
This is due to the small value of $\Delta A_\ell\approx1.6\times10^{-4}$ for this case.
We remark that this is only true for $\Delta\eta_\ell$, while the individual phase
shifts $\eta_\ell^i$ vary substantially (see example in Fig.~\ref{fig:eta_ell}) as they contain 
the term $\ell\frac\pi2$ and the nontrivial term $\ell(\ell+1)A^i_\ell$ in
Eq.~(\ref{eq:eta_ell-A_ell}) increases rapidly with $\ell$ as $A^i_\ell$
is not small (it contains the long-range contribution which is removed
in $\Delta A_\ell$). Again, as discussed above, the highly beneficial cancellations in obtaining 
$\Delta\eta_\ell$, namely the disappearance of the term $\ell\frac\pi2$ in Eq.~(\ref{eq:Delta-eta_ell})
and the subtraction of the long-range contributions of the $A^i_\ell$ terms in 
Eq.~(\ref{eq:DA_ell-exact}), lead to the
slowly varying value of $\Delta\eta_\ell$.
%(see Fig.~\ref{fig:Ai0-of-E}).
% avoid refering to Fig:Ai0-f-E, because it shows A_0 (not A_\ell)

%

\subsection{Locking and unlocking of phase shifts}
\label{sec:eta-lock}

The long-range cancellation encapsulated in
Eqs.~(\ref{eq:Delta-eta_ell-R}) and (\ref{eq:DA_ell-exact})  
has important consequences for the
phase shift difference $\Delta\eta_\ell(E)$ as a function of $E$ and
$\ell$.   As mentioned above, the phase shifts are locked with
respect to their $\ell$ dependence at low $\ell$, which is made clear in
Fig.~\ref{fig:Delta_eta_ell} where we see that $\Delta\eta_\ell$ varies little with 
$\ell$ for low partial waves; the phase locking region is identified by the shaded region.

\begin{figure}[t]
  \includegraphics[width=\linewidth] {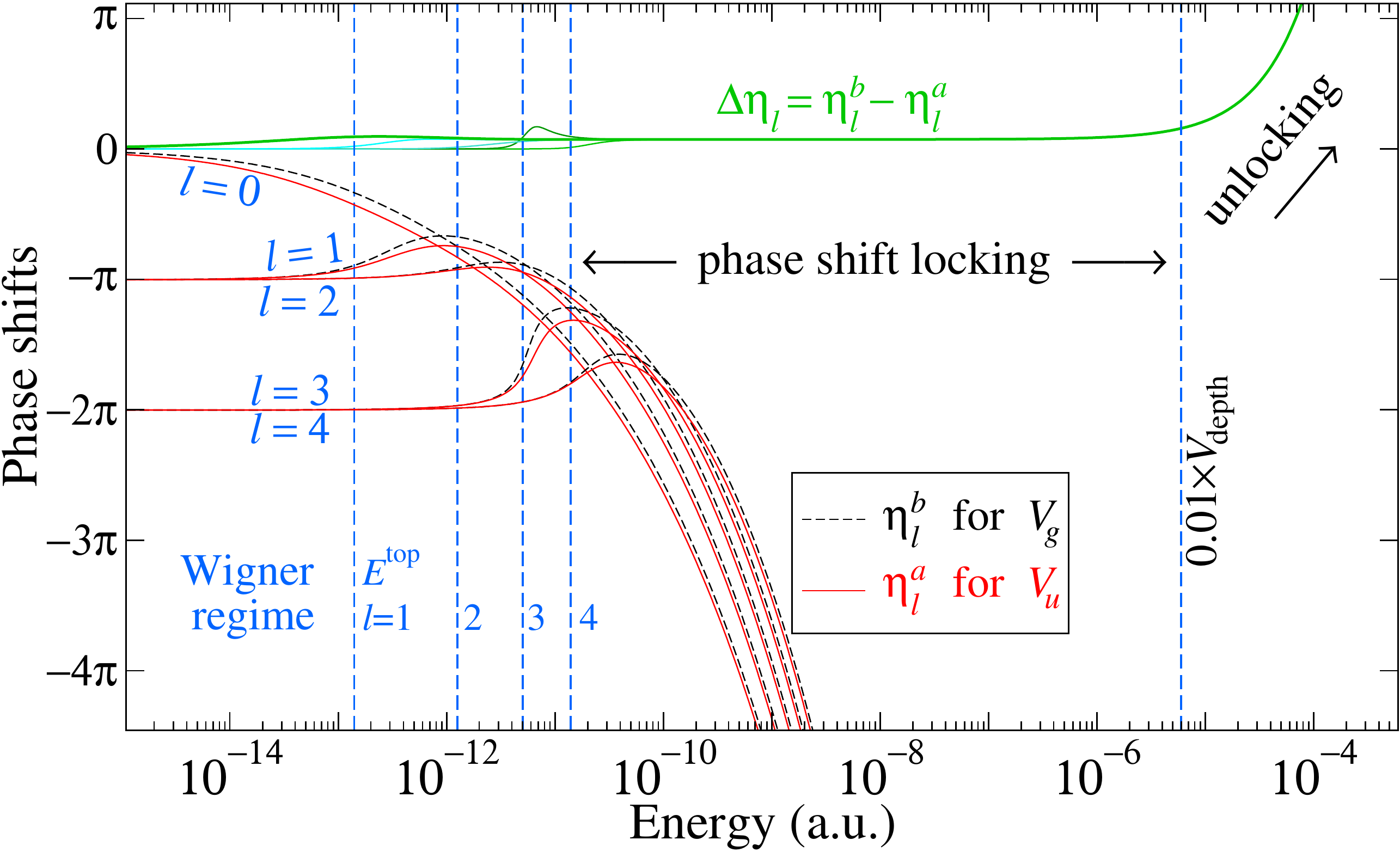}
  \caption{
\label{fig:eta-E}      
Energy dependence of the individual phase shifts $\eta_\ell^{a,b}$ and
their difference $\Delta\eta_\ell$
for \textsuperscript{170}Yb, with channels $a/b$ corresponding 
to $V_{u/g}$, respectively.  
The $\ell=1,2$ phase shifts are lowered by $\pi$ to account for the decrease 
of the number of bound states by one (Levinson's theorem); similarly, another 
bound state is lost for $\ell=3,4$ (see also steps in Fig.~\ref{fig:eta_ell}).
The vertical dashed lines indicate the heights of the centrifugal
barriers $E^\text{top}$ for $\ell=1,2,3,4$, and 
%($E^\text{top}=E_*$, $9E_*$, $36E_*$, and $100E_*$)
the boundary of the high-energy regime chosen at
$E=0.01\times V_\text{depth}$, where $V_\text{depth}$ is the depth of
the shallow potential $V_g$, which is a suitable  energy scale
at short-range (see Fig.~\ref{fig:Yb-potentials}); it separates the locking and unlocking regimes,
(see Sec.~\ref{sec:lock}).
  }
\end{figure}

Regarding the energy dependence, Fig.~\ref{fig:eta-E} shows
that the \emph{individual} phase shifts vary rapidly even at low $E$,
except for the Wigner regime of vanishing energies; it is also readily
apparent that $\eta^a_\ell(E)$ and $\eta^b_\ell(E)$ are \emph{locked}
to each other, which is due to the fact that the asymptotic tail of
the interaction dictates the $E$ dependence of $\eta_\ell^i(E)$ for
$E\ll V_\text{depth}$ and the two potential curves ($V_i$ with
$i=a,b$) are essentially identical at long-range (see Eq.~(\ref{eq:Va=Vb})).
%Figure~\ref{fig:eta-E} also shows
Consequently, the locking of the individual phase shifts ensures that the
% cancellation of their long-range contribution yields a
phase shift \emph{difference}, $\Delta\eta_\ell(E)$,
% that
is nearly constant across a wide range of energies, as illustrated in
Fig.~\ref{fig:eta-E}.

Figure~\ref{fig:eta-E} shows that $\Delta\eta_\ell(E)$ begins to vary
appreciably only when $E$ increases and becomes comparable to the
short-range energy scale $V_\text{depth}\sim 10^{-4}$ a.u.  In other words, the
individual phase shifts will \emph{unlock} gradually at higher
energies, as the short-range interaction begins to play a significant
role; this is expected because the only contribution to the integral
representation of the phase shift difference in
Eq.~(\ref{eq:Delta-eta_ell-R}) stems from the short-range region.
Similarly, for a fixed energy, the individual phase shifts will unlock
at high $\ell$ as $\Delta\eta_\ell$  begins to grow
(see Fig.~\ref{fig:Delta_eta_ell} and Sec.~\ref{sec:results}).

We emphasize that although the $E$ and
$\ell$ dependences of phase shifts are intertwined, their effects can
be analyzed separately, as we shall see in the next section. The locking and
unlocking of phase shifts dictates the behavior of the
resonant-exchange cross section.

%\end{figure*}

\section{The Quantal correction function}

\label{sec:F}

\subsection{General formulation}
\label{sec:F-general}

We first recall that Eq.~(\ref{eq:csec-def-approx}) truncates the
partial-wave expansion at $\ell\approx L$ and that $\Delta\eta_\ell$ 
varies slowly for $\ell<L$ (see Fig.~\ref{fig:Delta_eta_ell});
consequently the partial-wave terms of the cross section in
Eq.~(\ref{eq:csec-def}) vary slowly with $\ell$. This justifies considering 
$\ell$ as a continuous variable and
converting the sum into an integral:
\[
\sum_{\ell=0}^L  (2\ell+1)\sin^2[\Delta\eta_\ell]\;\Rightarrow
\int_0^L  \!d\ell\,(2\ell+1)\sin^2[\Delta\eta(\ell)].
\]
Note that the integration renders $\ell=0$ a point of measure zero,
which would have the effect of discarding the $s$-wave term; the
remedy is to simply add the $s$-wave term separately.  Hence, the
correct conversion of the truncated partial-wave sum into an integral
reads
\begin{equation}
 \label{eq:integral-lambda}
   \sum_{\ell=0}^{L} (2\ell+1) \sin^2[\Delta\eta_\ell] \approx  \sin^2[\Delta\eta_0]
   +\! \int_0^\Lambda \!\!\!d\lambda \, \sin^2 [\Delta\eta(\lambda)],
\end{equation}
where $\lambda\equiv\ell(\ell+1)$ is a newly introduced integration
variable, and $\Lambda = L(L+1)$ is the energy-dependent cutoff which
was already defined in Eq.~(\ref{eq:LL4}).  We note that according to Eq.~(\ref{eq:delta_0}),
\begin{equation}   
   \Delta\eta_0 (k)=  \Delta N_0 \pi + \Delta\delta_0 (k) \; ,
   \label{eq:Delta-eta-delta_0}
\end{equation}
where $\Delta N_0 =N_0^a -N_0^b$ is an integer and $\Delta\delta_0 =\delta_0^a -\delta_0^b$, so that
$\sin^2[\Delta\eta_0] = \sin^2[\Delta\delta_0]$.

Substituting the
expressions above in Eq.~(\ref{eq:csec-def}) yields an approximation
for the cross section,
\begin{equation}   
  \sigma_{\rm exc}(E) \approx
  \sigma_0(E) + F_\Lambda(E)\sigma_L (E),
     \label{eq:sigma=sigma_0+Fsigma_L}
\end{equation}
where $\sigma_0=\frac\pi{k^2}\sin^2[\Delta\delta_0]$  is the $s$-wave
cross section, $\sigma_L=\frac\pi{k^2}\Lambda$ is the Langevin cross
section, and $F_\Lambda$ is defined as
\begin{equation}
   \label{eq:F-Lambda}
  \begin{split}
    F_\Lambda(E)
 & \equiv \frac{1}{\Lambda(E)} \int_0^{\Lambda(E)} \!\!\!
     d\lambda \, \sin^2 [\Delta\eta(E;\lambda)].
  \end{split}
\end{equation}
This \emph{approximate} version of the quantal correction function
obeys $F_\Lambda \leq 1$, which justifies the interpretation of
$\sigma_L$ as an upper limit of the quantal cross section at higher
energies where the $s$-wave contribution becomes negligible and we
thus have $\sigma_\text{exc}\approx
F_\Lambda\sigma_L\lesssim\sigma_L$.  Moreover, at sufficiently high
energies it may become justified to replace $\sin^2[\cdots]$ by its
average value of $\frac1{2}$ in Eq.~(\ref{eq:F-Lambda}), yielding the
high energy limit $F_\Lambda\to\frac1{2}$ and hence
$\sigma_\text{exc}\to{1\over2}\sigma_L$.

Regarding the approximations employed so far, we expect the
conversion of the partial-wave sum into an integral to be adequate
at high energies where many partial waves contribute; conversely, it
becomes inadequate at ultralow energies (near the Wigner
regime) where only a few partial waves are significant.  We also note
that due to the truncation of high partial waves ($\ell>L$), the
contribution of most shape resonances is lost, except for resonances
which occur for $\ell\lessapprox L$\@.  The latter
are still included in Eq.~(\ref{eq:F-Lambda}), albeit in an altered
fashion due to the conversion of the sum into an integral, leading to 
an incomplete account of shape resonances. 
We will instead employ a simplifying approximation for the
$\lambda$-dependence of $\Delta\eta(\lambda)$ in
Eq.~(\ref{eq:F-Lambda}) which will eliminate all
shape resonances and, more importantly, will allow us to obtain a
closed-form expression for the energy dependent \emph{background} of
the quantal correction function.

\begin{figure}[t]
  \includegraphics[width=0.95\linewidth]
                  {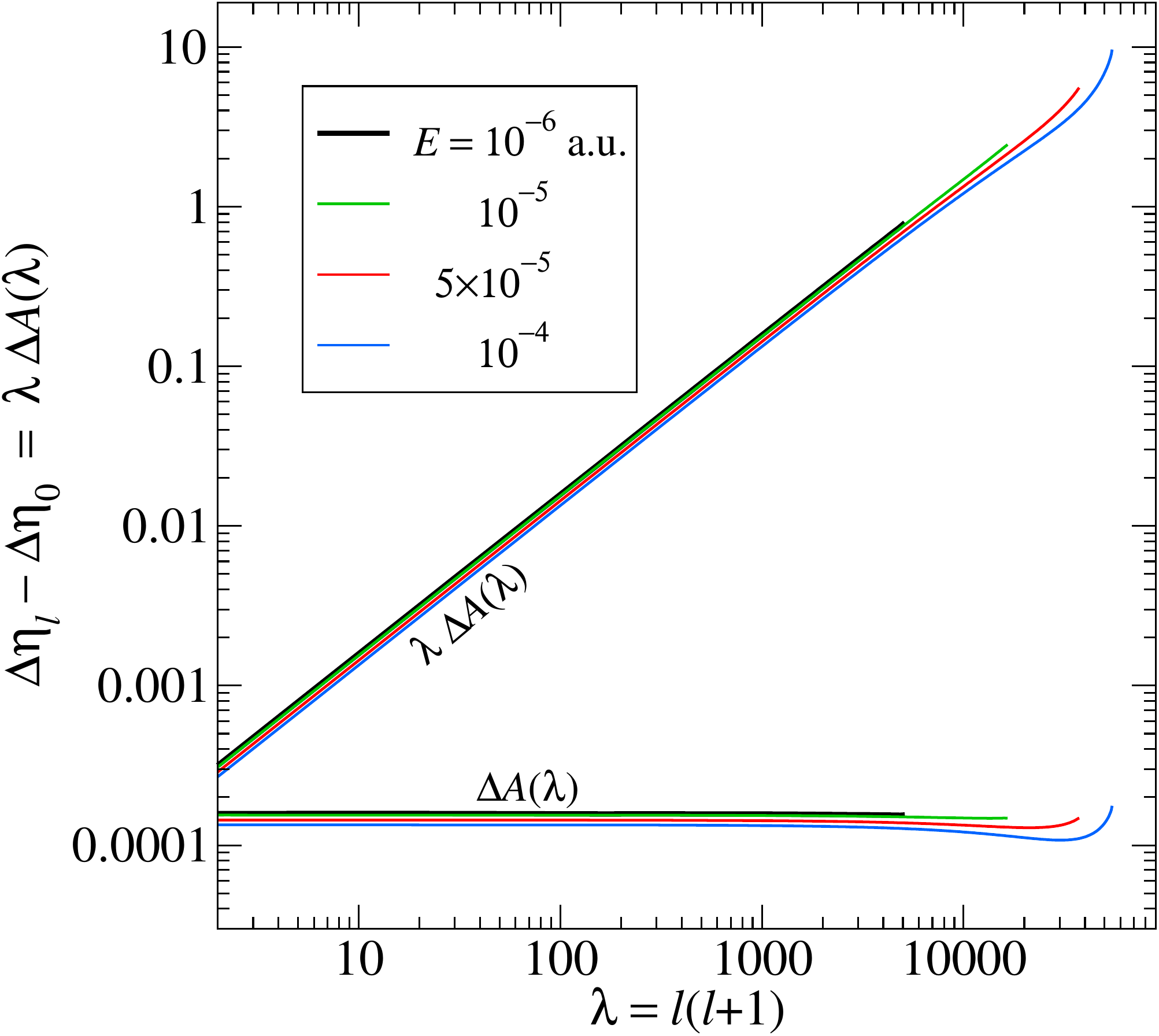}
  % log-log plot vs lambda, for several E=1e-8, 1e-6, 1e-5, x?e-5, 1e-4
                  \caption{
                    \label{fig:DA-ell}
                    Partial-wave dependence of the phase-shift difference for
  different values of the scattering energy (see legend).  The data
  set for each energy, $E$, is truncated at the cutoff $\Lambda(E)$.
  As in Fig.~\ref{fig:Delta_eta_ell}, the results are nearly identical
  for all isotopes (due to the fact that the $s$-wave term was
  subtracted out).
 %  shown here for  isotope  $^{172}\text{Yb}$
}

\end{figure}

\subsection{Expression of the quantal correction function}

\label{sec:F-A}

Although $F_\Lambda$ is an approximate version of the quantal
correction function ${\cal F}$ appearing in Eq.~(\ref{eq:F-exact-intro}), 
the integrand in Eq.~(\ref{eq:F-Lambda}) still
contains the exact $\lambda$ dependence of $\Delta\eta(\lambda)$.  The
latter was given in terms of $\ell$ and $\Delta A_\ell$ in
Eq.~(\ref{eq:Delta-eta_ell}) which we recast here as a function of
$\lambda=\ell(\ell+1)$,
\[
    \Delta\eta(\lambda) =
    \Delta\eta_0 - \lambda\cdot\Delta A(\lambda),
\]
and thus, Eq.~(\ref{eq:F-Lambda}) reads
\begin{equation}
  F_\Lambda  =  \frac{1}{\Lambda} \int_0^{\Lambda} \!  d\lambda \,
  \sin^2 [\Delta\eta_0 - \lambda\cdot\Delta A(\lambda)] \;.
  \label{eq:F-Lambda-A} 
\end{equation}
We now employ one last approximation in the equation above.
Specifically, we shall use $\Delta A(\lambda)\approx\Delta
A_0\equiv\lim_{\ell\to0}\Delta A_\ell$, which is justified at low
energies;  indeed, for $E\ll V_\text{depth}$,
$\Delta A(\lambda)$ is nearly constant for
$\lambda<\Lambda(E)$, as shown in Fig.~\ref{fig:DA-ell}.  Accordingly,
the phase shift difference is approximated as a linear function of
$\lambda$, i.e., quadratic in $\ell$,
\begin{equation} 
  \label{eq:Delta-eta_0}
    \Delta\eta(\lambda)
     \approx  \Delta\eta_0 - \lambda \cdot\Delta A_0 
     =  \Delta\eta_0 - \ell(\ell+1)\cdot \Delta A_0.
\end{equation}
This approximation is compared with the exact results in
Fig.~\ref{fig:Delta_eta_ell}, which shows good agreement.

Finally, the linearization of $\Delta\eta(\lambda)$ allows for the
integration in Eq.~(\ref{eq:F-Lambda}) to be done analytically. 
Using Eq.(\ref{eq:Delta-eta-delta_0}) so that 
$\sin^2 [\Delta\eta_0 - \lambda\cdot\Delta A(\lambda)] = \sin^2 [\Delta\delta_0 - \lambda\cdot\Delta A(\lambda)]$,
we obtain a closed-form (denoted $F$) for the quantal correction
function,
\begin{eqnarray}
   \mathcal F \approx
  F & \equiv &
  \frac{1}{\Lambda} \int_0^\Lambda d\lambda \sin^2(\Delta\delta_0- \lambda\cdot \Delta A_0),
  \nonumber \\
 & = &
 \frac{1}{2} - \frac{\sin( \Lambda\cdot\Delta A_0)}{2 \Lambda\cdot\Delta A_0}
 \cos(2\Delta\delta_0 -  \Lambda\cdot\Delta A_0).
  \label{eq:F}
\end{eqnarray}
Accordingly, the cross section  reads
\begin{equation}   
  \sigma_{\rm exc}(E) \approx
  \sigma_0(E) + F(E)\sigma_L (E).
   \label{eq:sigma=sigma_0+Fsigma_L}
\end{equation}
The $s$-wave contribution ($\sigma_0$) is dominant in the zero energy
limit and ensures the correct behavior of $\sigma_\text{exc}(E)$ in
the Wigner regime, provided that the $s$-wave phase shifts
$\eta_0^{a,b}$ are computed quantum mechanically.  We also remark that
$s$-wave scattering plays an additional role in the quantal correction
function, which in turn accounts for the contribution of higher
partial waves; namely, the term $2\Delta\delta_0(E)$ in Eq.~(\ref{eq:F})
has a strong effect on $F(E)$ and hence $\sigma_\text{exc}(E)$.
Indeed, our results will show that $s$-wave scattering effects persist
to energies that are many orders of magnitude above the $s$-wave
regime, and in fact dictate the behavior of $\sigma_\text{exc}(E)$ 
over a broad range of energies.

\subsection{Low and high energy  limits}
\label{sec:F-high-low}

In the limits of low or high energy, our new result~(\ref{eq:F})
yields much simpler expressions for the quantal correction function.
First, when $E\to0$ we have $\Lambda(E)\to0$ ensuring $\Lambda\Delta
A_0 \ll 1$, and we recover the low-energy approximation obtained in
our previous work~\cite{PRL-2018-Simbotin-Cote},
$F\approx\frac1{2}(1-\cos2\Delta\delta_0)=\sin^2(\Delta\delta_0)$.  In the
opposite extreme regime, i.e., at high energies, $\Lambda(E)$
increases considerably and we have $\Lambda\cdot\Delta A_0 \gg 1$, yielding
the high energy limit $F\to\frac1{2}$ which is approached in an
oscillatory fashion with an amplitude diminishing as
$\frac1{\Lambda(E)}\sim E^{\frac{2-n}n}$.  Hence, in the limits of low
and high energy, respectively, the quantal correction
function~(\ref{eq:F}) reduces to
\begin{equation}
 F \to  \left\{ \begin{array}{ll}
   \sin^2 \Delta\delta_0
   & \text{ for}\quad \Lambda\cdot\Delta A_0 \ll 1 \quad (\text{low } E) 
   \\
  %\displaystyle
   \tfrac{1}{2} + \genfrac(){0pt}{1}{\text{small}}{\text{osc.}}
   & \text{ for}\quad \Lambda\cdot\Delta A_0 \gg 1 \quad (\text{high } E),
      \end{array} \right. 
\label{eq:F-limits}
\end{equation}
and similarly for the cross section we have
\begin{equation}
 \sigma_\text{exc} \to  \left\{ \begin{array}{ll}
   \left(\frac\pi{k^2}+\sigma_L\right) \sin^2 \Delta\delta_0
   &  \quad (\text{low } E) 
   \\
  %\displaystyle
   \tfrac{1}{2}\;\sigma_L + \genfrac(){0pt}{1}{\text{small}}{\text{osc.}}
   &  \quad (\text{high } E).
      \end{array} \right. 
\label{eq:sigma-limits}
\end{equation}

Finally, we note that for long-range %inverse power-law 
potentials of the form $r^{-n}$ with $n>3$, the phase shift $\delta_0^i \sim -ka_i$ as $k\rightarrow 0$ (or $E\rightarrow 0$), where $a_i$ is the scattering length corresponding to potential $V_i$
\cite{Harald-book,Cote-thesis,Gribakin-1993}. In that case, the quantal correction function takes the simple form
$F\approx \sin^2 \Delta\delta_0 = k^2 (\Delta a)^2$, where $\Delta a = a_a - a_b$, and the cross section
$\sigma_\text{exc} \approx \pi (a_a - a_b)^2$.

 The validity of our general expression~(\ref{eq:F}) and its simpler 
 forms above will be discussed in the next section
 where we compare them with the exact quantal results.

%  \newpage

\section{Results and discussion}
\label{sec:results}

We now give a detailed presentation of our computed cross sections for
charge-transfer collisions of Yb$^+$ with its neutral parent atom.  We
study this ion-atom system as an illustrative example of the more
general case of resonant-exchange scattering, and we recall that the
theoretical framework in the previous sections is indeed applicable to
any system which has interactions behaving asymptotically as an 
inverse power, i.e., $V(r)\approx-C_n/r^n$ for $r\to\infty$ for any
$n>2$.

\subsection{Overview}
\label{sec:results-overview}

%\newpage

\begin{figure}[t]
   \includegraphics[width=0.99\linewidth]
                   {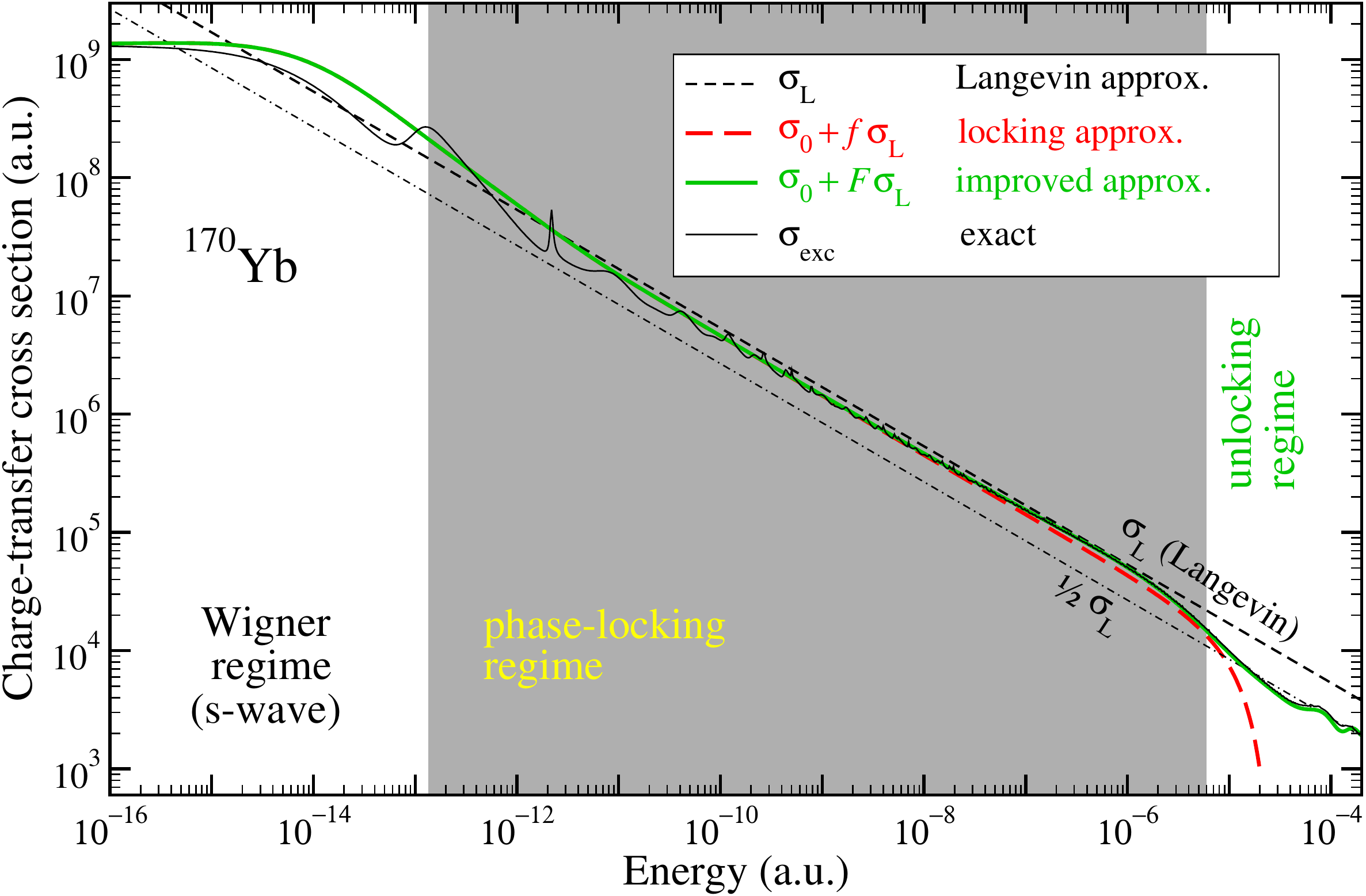}
   \includegraphics[width=0.99\linewidth]
                   {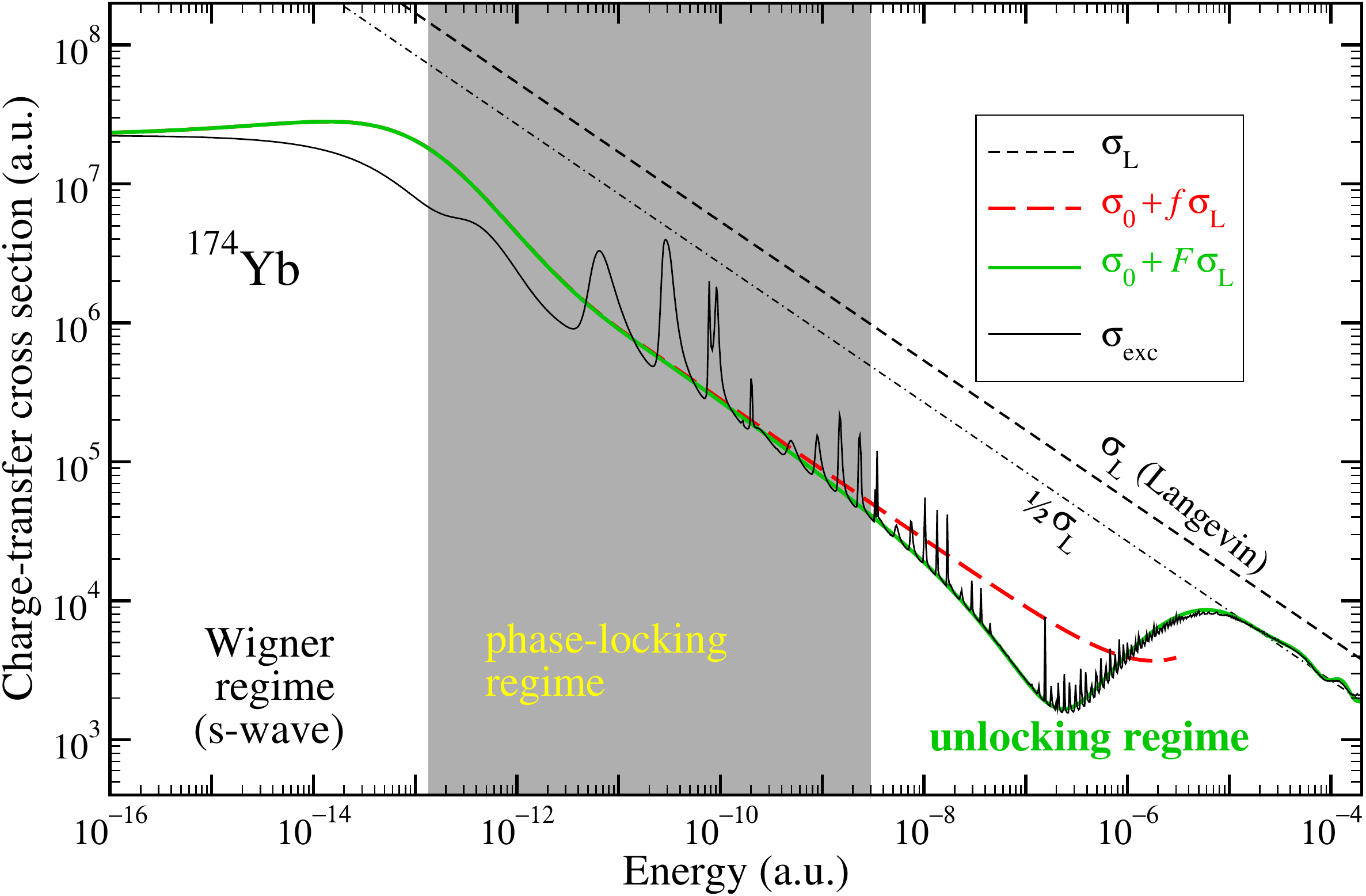}
\caption{\label{fig:sigma-regimes}
  The charge-transfer cross section for isotopes $^{170}$Yb (top) and
  $^{174}$Yb (bottom).  The thin black line is for the quantal (exact)
  result, while the green curve is our newly obtained approximation
  using the quantal correction function $F(E)$ in Eq.~(\ref{eq:F}).
  Also shown are the locking approximation (thick red dash line) which
  uses the much simpler version of the quantal correction function,
  $f(E)$ in Eq.~(\ref{eq:f-lock}), and the Langevin cross section
  $\sigma_L$ (black dash line) together with $\frac{1}{2}\sigma_L$
  (thin dot-dash line).  The latter represents the correct high-energy
  limit, which the quantal result approaches in an oscillatory
  fashion.  The three separate regimes discussed in the main text are
  labeled accordingly; for emphasis the shaded area covers the
  intermediate regime where phase shift locking is in effect.
  }
\end{figure}

% We first recall that
An overview of our computed (quantal) cross
sections $\sigma_{\rm exc}$ and the corresponding quantal correction 
functions ${\cal F}$ is shown in Fig.~\ref{fig:Yb-sigmas-F-ratio}
% and~\ref{fig:F-exact-semilog},
where three distinct regimes can be identified as the main features of
the energy dependence.  These energy regimes are illustrate more
clearly in Fig.~\ref{fig:sigma-regimes} 
for isotopes $^{170}\text{Yb}$ and $^{174}\text{Yb}$.  
At vanishingly low energies, we encounter the Wigner regime
where the $s$-wave contribution, $\sigma_0(E)$, dominates and
approaches a constant value $\pi (a_u-a_g)^2$ as $E\to0$ [$a_{g/u}$: scattering 
lengths corresponding to $V_{g/u}$], while the Langevin cross
section $\sigma_L$ diverges as $E^{-1/2}$.  Next we have a regime of intermediate energies
where $\sigma_\text{exc}(E)$ exhibits a different behavior
for each isotope (see also Fig.~\ref{fig:Yb-sigmas-F-ratio}).
While the quantal cross sections $\sigma_\text{exc}(E)$ has roughly the $E^{-1/2}$ dependence
of the Langevin cross section $\sigma_L(E)$, their magnitudes usually differ,  
except for a few specific isotopes.
%
%Consequently, $\sigma_L(E)$ is in stark disagreement with the quantal
%cross section, except for some isotopes which show fair agreement; the
%comparison will be further discussed in the subsequent sections.  
We
remark that the domain of intermediate energies extends across many
orders of magnitude and corresponds to the regime of phase shift
locking, which was discussed in Sec.~\ref{sec:eta-lock}; thus, we 
refer to it as the locking regime, as indicated in
Fig.~\ref{fig:sigma-regimes}.  Finally, we have the high energy regime
where the cross sections for all isotopes begin to coalesce together,
irrespective of their vastly different behavior at intermediate
energies.  Specifically, as the energy increases, $\sigma_\text{exc}$
attains a more uniform behavior with small oscillations about an
average value of $\frac1{2}\sigma_L$.  This was also shown very
clearly in Fig.~\ref{fig:F-exact-semilog} where one can see that the
quantal correction functions for all isotopes approach the high-energy
limit, $\mathcal F\to\frac1{2}$, in an oscillatory fashion.  This
behavior was obtained as the high-energy limit (see
Eqs.\eqref{eq:F-limits} and \eqref{eq:sigma-limits}) of the general
expression~\eqref{eq:F}.  The high energy regime corresponds to the
unlocking of phase shifts, as discussed in Sec.~\ref{sec:eta-lock},
and is dubbed the unlocking regime.

In the next sections, we pay special attention to the locking and unlocking regimes.
We recall that the phase shifts $\eta^{a,b}_\ell(E)$ are locked
together at low $E$ and low $\ell$, and subsequently they begin to
unlock as $E$ and $\ell$ increase, as explained in
Sec.~\ref{sec:eta-lock}\@.  This has important consequences for the
phase shift difference, $\Delta\eta_\ell$, and thus for the resonant-exchange
cross section.  Indeed, the key ingredients of our 
expression~(\ref{eq:F}) are the locking and unlocking of phase shifts
for high partial waves---together with the $s$-wave phase shift
difference---%
%$\Delta\eta_0$ itself, 
and their interplay is responsible for the range of behaviors
shown by the cross sections for different isotopes.

\subsection{Phase locking regime }
\label{sec:lock}

\begin{figure}[h]
  \includegraphics[width=1.0\linewidth]
                   {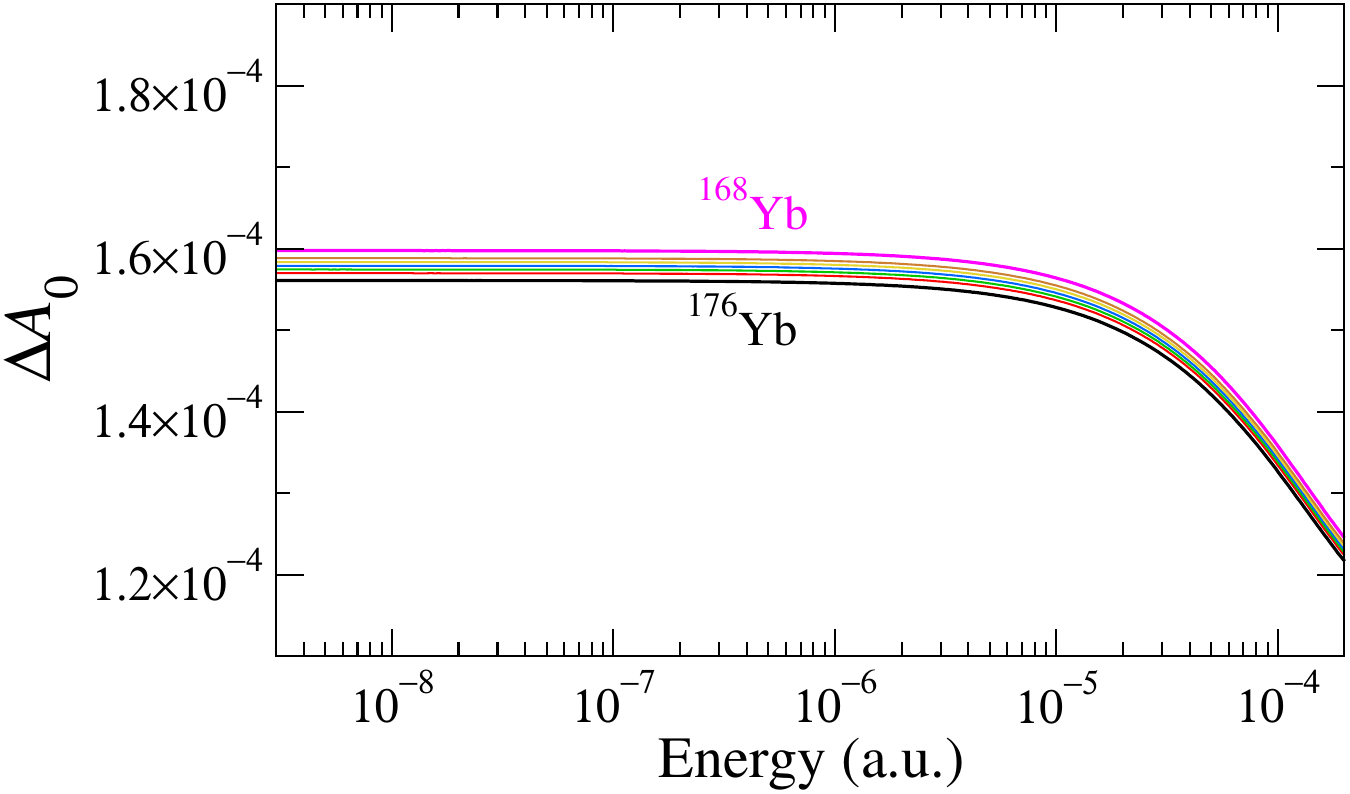}
\caption{\label{fig:DA0} Energy dependence of $\Delta A_0(E)$ for all
  isotopes of Yb.  The coefficient $\Delta A _0$
  changes very little with the isotopes because the mass differences are relatively small
  for heavy systems (such as Yb).  Moreover, the simple mass scaling $\Delta
  A_0\propto \mu^{-1/2}$ \cite{PRL-2018-Simbotin-Cote} decreases monotonically with the isotope mass; 
  hence, the seven curves shown here are ordered
  accordingly from the lightest (\yb{168}, top) to the heaviest
  isotope (\yb{176}, bottom).}
\end{figure}

The approximation underpinning the regime of phase-shift locking, introduced 
in our previous work \cite{PRL-2018-Simbotin-Cote}, consists in neglecting the
$\ell$-dependence of the phase-shift \emph{difference}, 
$\Delta\eta_\ell(E)\approx\Delta\eta_0(E)$, or
\begin{equation}
  \label{eq:Delta-eta-lock}
%\Delta\eta(\lambda;E)=\Delta\eta_\ell(E)\approx\Delta\eta_0(E).
   \Delta\eta_\ell(E)\approx\Delta\delta_0(E) \;,
\end{equation}
where we omit $\Delta N_0 \pi$ in Eq.~(\ref{eq:Delta-eta-delta_0})   
since it does not contribute to $\sin^2 [\Delta\eta_0]$.
This is a simplified form of Eq.~(\ref{eq:Delta-eta_0}) and is valid
provided that the term $\lambda\cdot\Delta A_0=\ell(\ell+1)\cdot\Delta A_0$
remains negligible, a requirement that can be expressed as a
strong inequality for $\ell$ or $\lambda$,
\[
\lambda = \ell(\ell+1)
\ll \left|\frac{\Delta\delta_0}{\Delta A_0}\right|,
\]
and will now be used to estimate the extent of the locking regime.
Specifically, for our illustrative example we use the value $\Delta
A_0\approx 1.6\times10^{-4}$ (see Fig.~\ref{fig:DA0}) and we assume
$\Delta\delta_0$ is of the order unity, yielding
$\frac{\Delta\delta_0}{\Delta A_0}\approx6.25\times10^3$.  

This large
value will ensure the validity of the locking approximation up to
fairly high partial waves and up to energies high above the
$s$-wave regime; indeed, if we consider one order of magnitude as a
sufficient threshold for neglecting the partial-wave dependence in
Eq.~(\ref{eq:Delta-eta_0}), i.e., $\lambda\cdot\Delta A_0
\lesssim0.1\ll1\approx|\Delta\delta_0|$, then the strong
inequality above will be fulfilled for $\lambda\lesssim 625$ and hence
$\sqrt\lambda\approx\ell\lesssim 25$.  This upper limit for the
partial waves can now be used to find the extent of the locking regime
in the energy domain; namely, the energy-dependent partial-wave cutoff
$\Lambda(E)=L(L+1)$ defined in Eq.~(\ref{eq:LL4}) must obey the
restriction $L(E)\lesssim25$ or $\Lambda(E)\lesssim625$, and we invert
the relationship~(\ref{eq:LL4}) to find $E(\Lambda)$ which yields an
upper limit for the energy: $E\lesssim E(625)\approx10^5E_*$.  Hence,
phase shift locking can persist up to energies which are many orders
of magnitude above the $s$-wave Wigner regime ($E\lesssim E_*$).
Finally, using the value of the energy scale,
$E_*=1.4\times10^{-13}$~a.u. (see Sec.~\ref{sec:theory-example}), the upper limit of the locking regime
can be converted to atomic units: $E\lesssim1.4\times10^{-8}$~a.u.

However, the actual extent of the locking regime will depend
critically on the value of $\Delta\delta_0$ (which was assumed to be of
the order unity, but can differ greatly for each isotope).  Indeed, as
illustrated in Fig.~\ref{fig:sigma-regimes}, the locking regimes for
different isotopes can span vastly different energy domains. Moreover,
the energy dependence of the cross section can also change drastically
with the isotope.  These important aspects of the isotopic dependence
will be discussed in the next two subsections.

Although the locking approximation~(\ref{eq:Delta-eta-lock}) seems 
drastic, it yields a nontrivial expression for the cross section;
indeed, we first write the simplified form of the quantal correction
function in the locking approximation, denoted $f$:
\begin{equation}
  \label{eq:f-lock}
 f(E) = \sin^2[\Delta\delta_0(E)],
\end{equation}
which is readily obtained by using Eq.~(\ref{eq:Delta-eta-lock})
directly in Eq.~(\ref{eq:F-Lambda-A}).  Note that $f(E)$ is consistent
with our improved approximation, $F(E)$, in Eq.~(\ref{eq:F}).  Namely,
$f(E)$  is precisely the low energy limit of
$F(E)$ obtained in Eq.~(\ref{eq:F-limits}).
Accordingly, the locking approximation for the cross section yields
\begin{equation}\label{eq:sigma-lock}
  \begin{split}
\sigma_\text{exc}\approx    \sigma^\text{lock}_{\rm exc}(E)
    &\equiv \sigma_0(E) + f(E) \sigma_L (E)
    \\
    &= \left( \frac\pi{k^2} + \sigma_L(E) \right)  \sin^2 \Delta\delta_0 (E),
  \end{split}
\end{equation}
which is precisely the low-energy limit~\eqref{eq:sigma-limits} and is
in good agreement with the exact results at low energies.  As shown in
our previous work~\cite{PRL-2018-Simbotin-Cote}, and also illustrated here in
Fig.~\ref{fig:sigma-regimes}, the locking
approximation~\eqref{eq:sigma-lock} is indeed valid across the Wigner
and locking regimes.
%In our previous work~\cite{prl-s-wave} we surveyed a wide range of
%resonant-exchange scattering systems
%and we found good agreement for $\sigma^{\rm lock}_{\rm exc}(E)$
%and the exact results at ultralow energies;
However, the locking approximation becomes inadequate at high
energies, as the phase shifts begin to unlock.  As shown in
Fig.~\ref{fig:sigma-regimes}, our improved approximation~\eqref{eq:F}
for the quantal correction function extends the agreement with the
exact results to higher energies by taking into account the unlocking
of phase shifts (see discussion in Sec.~\ref{sec:unlock}).

We also remark that although the contribution of higher partial
waves grows quickly as the energy increases above the Wigner domain
and becomes dominant in the locking regime, the behavior of the cross
section is still driven by the $s$-wave phase shift difference
($\Delta\delta_0$) which is prominent in the equation above.  Indeed, as
shown in our previous work~\cite{PRL-2018-Simbotin-Cote}, the signature of s-wave
scattering persists to energies very high above the Wigner regime.

Finally, we note that $\sigma_{\rm exe}$ obtained using Eq.~(\ref{eq:F}) 
over-estimates the cross section compared to the exact quantal value
at energies around $E_*$, where $\ell=1$ kicks in (see Fig.~\ref{fig:sigma-regimes}). 
This is simply due to the inadequacy of replacing the partial-wave sum by
integration over $\ell$ when there are only a few terms
contributing.  However, the agreement improves rapidly as $E$
grows. Moreover, the over-estimate gradually disappears as $E$
decreases into the Wigner regime, since $\delta_0^i\approx-ka_i$
(for $n>3$) so that $\sin^2\Delta \delta_0 \sim E (\Delta a)^2$; then
$\sigma_0\sim k^{-2}\sin^2\Delta\delta_0\sim\text{const}$ while
$\sigma_L \sim E^{-\frac{2}{n}}$ leads to
$\sigma_L\sin^2\Delta \delta_0 \sim E^{-\frac{2}{n}+1} = E^{\frac{n-2}{n}}$.
For $n=4$, the threshold behavior for these two terms are 
a constant and $E^{1/2}$, respectively, with the second term vanishing as $E\rightarrow 0$.

\begin{figure}[b]
  \includegraphics[width=1.0\linewidth]
                   {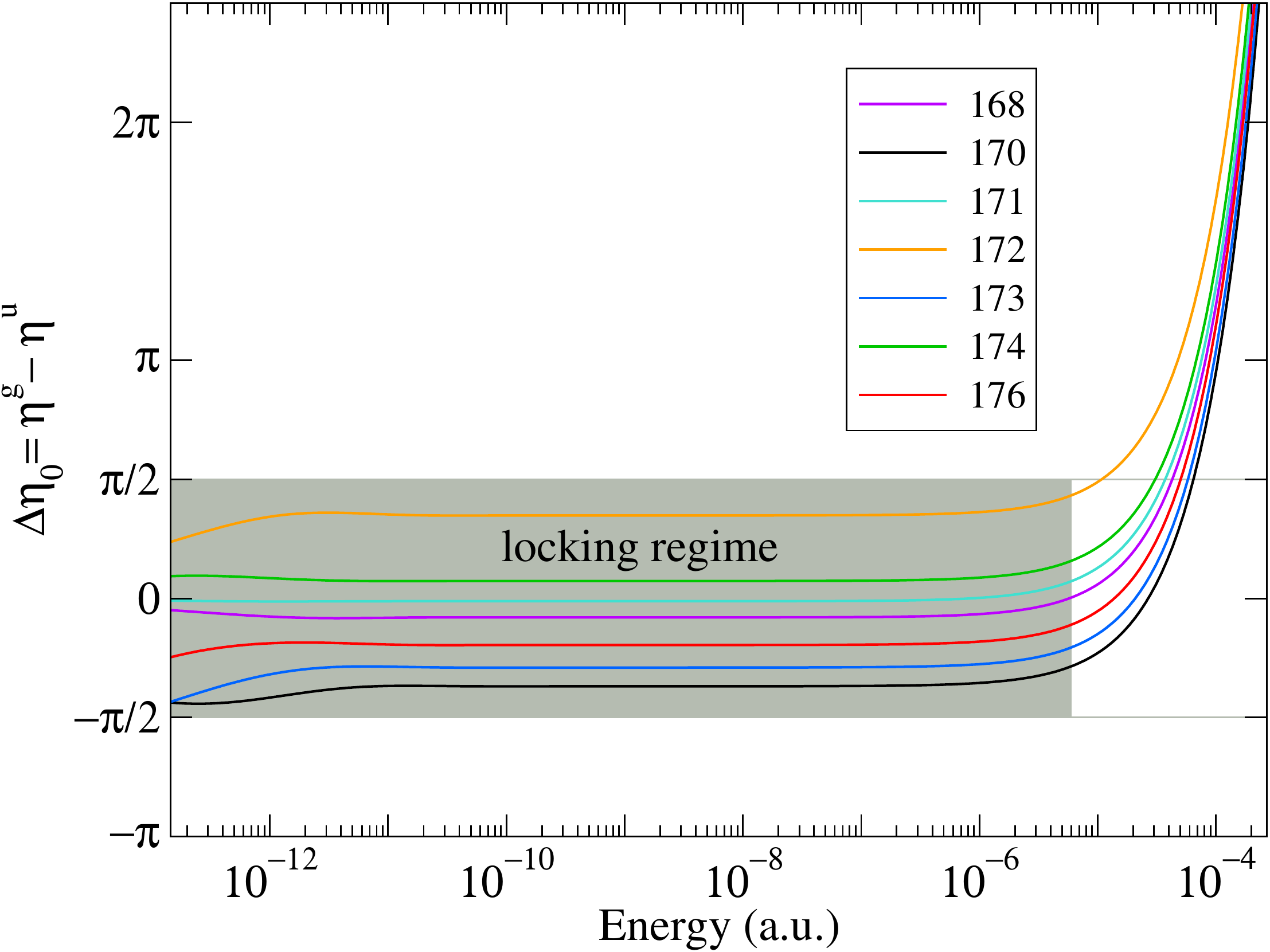}
                   \caption{\label{fig:Deta-lock}
                     The phase shift difference  $\Delta\eta_0$ varies
                     considerably between isotopes 
                     (unlike $\Delta A_0$, which is nearly the same for
                     all isotopes, as shown in Fig.~\ref{fig:DA0}).
}
\end{figure}

%\newpage

\subsection{%Accidental suppression
             Ubiquity of suppression
            % Suppression
             and enhancement inside the locking regime}

In the previous section, we estimated the extent of the locking regime
under the assumption that the value of $\Delta\delta_0$ at low and
ultralow (but not vanishing) energies is of the order unity;
consequently, the upper boundary of the locking regime was determined
by the value of $\Delta A_0$.  However, Fig.~\ref{fig:DA0} shows that
the coefficient $\Delta A_0$ is nearly the same for all isotopes,
while the locking regime can be quite different for each isotope (see
Fig.~\ref{fig:sigma-regimes}).  This is due to the fact that, although
the phase shift difference $\Delta\delta_0$ has practically the same energy dependence
for all isotopes throughout the locking regime and also at higher
energies, the values of $\Delta\delta_0$ for different isotopes are
shifted by large amounts relative to each other, as shown in
Fig.~\ref{fig:Deta-lock}.  Hence, the value of $\Delta\delta_0$ inside
the locking regime can vary widely from one isotope to the next;
namely, $\Delta\delta_0$ can be of the order unity or even approaching
$\pm\frac\pi2$, or it can be vanishingly small. The variability of $\Delta\delta_0$
is connected to the sensitivity of the position of the last bound 
level to the reduced mass in the respective potential wells $V_{g/u}$, 
itself linked to large variations in the scattering lengths for different
isotopes \cite{Harald-book,Cote-thesis,Gribakin-1993}. These ``accidental''
occurrences have important consequences, as we discuss next.

We recall that $\Delta\delta_0 (E)$ for a given isotope is nearly
constant in the locking regime, as discussed in
Sec.~\ref{sec:eta-lock} and as shown in Fig.~\ref{fig:Deta-lock}.
For cases with a vanishingly small
phase shift difference $\Delta\delta_0\approx0$, 
the quantal correction function given by Eq.~(\ref{eq:f-lock}) has 
exceedingly low values,  i.e. $f=\sin^2\Delta\delta_0\ll1$. Hence
Eq.~\eqref{eq:sigma-lock} will yield a markedly {\it suppressed} cross
section, i.e., $\sigma_\text{exc}\ll\sigma_L$ inside the locking
regime; such suppression cases are abundantly clear in our results,
e.g., isotopes $^{168}$Yb, \yb{171}, and \textsuperscript{174}Yb 
(see Figs.~\ref{fig:Yb-sigmas-F-ratio} and \ref{fig:F-exact-semilog}).
%% ....Note also that the exact cross section is strikingly different
%% than the Langevin cross section;
Moreover, the extent of the locking regime is diminished in suppressed
cases; indeed, when $\Delta\delta_0$ is small, the $\Lambda$ dependence
in Eq.~(\ref{eq:F}) rapidly becomes significant and the locking
approximation $f(E)=\sin^2\Delta\delta_0(E)$ ceases to be valid;
it must then be replaced by $F(E)$, i.e., the improved
approximation~\eqref{eq:F} of the quantal correction function. 
We note that this case results from the scattering lengths  $a_i$ of the 
two potential curves $V_i$ being (accidentally) nearly equal, so that the $s$-wave 
phase-shift \emph{difference} $\Delta\delta_0(E)$ is vanishingly small and remains so 
across a domain of energies much wider than the Wigner regime.

Another type of behavior takes place for cases when $\Delta\delta_0\approx\pm\frac\pi2$. 
As opposed to the suppressed quantal correction function, we obtain values reaching the
maximum of unity inside the locking regime, with
$f=\sin^2\Delta\delta_0\approx 1$. The cross section then approaches
its upper bound, i.e., $\sigma_\text{exc}\approx\sigma_L$, and we refer to these as
{\it enhanced} cases. The isotopes $^{170}$Yb and \yb{172} are examples of enhanced
cases
(see Figs.~\ref{fig:Yb-sigmas-F-ratio} and \ref{fig:F-exact-semilog}).
Finally, in between suppression and
enhancement we have the third case which we simply dub the {\it average} case
corresponding to basically substituting $\sin^2\Delta\delta_0$ by its ``average" of
$\frac{1}{2}$, or $\Delta\delta_0\approx\pm\frac\pi4$, 
which yields $f\approx\frac12$ and hence $\sigma_\text{exc}\approx\frac12\sigma_L$.
The remaining isotopes $^{173}$Yb and \yb{176} illustrate this case (see Figs.~\ref{fig:Yb-sigmas-F-ratio} and \ref{fig:F-exact-semilog}).

\begin{figure}[b]
  \includegraphics[width=0.9\linewidth]{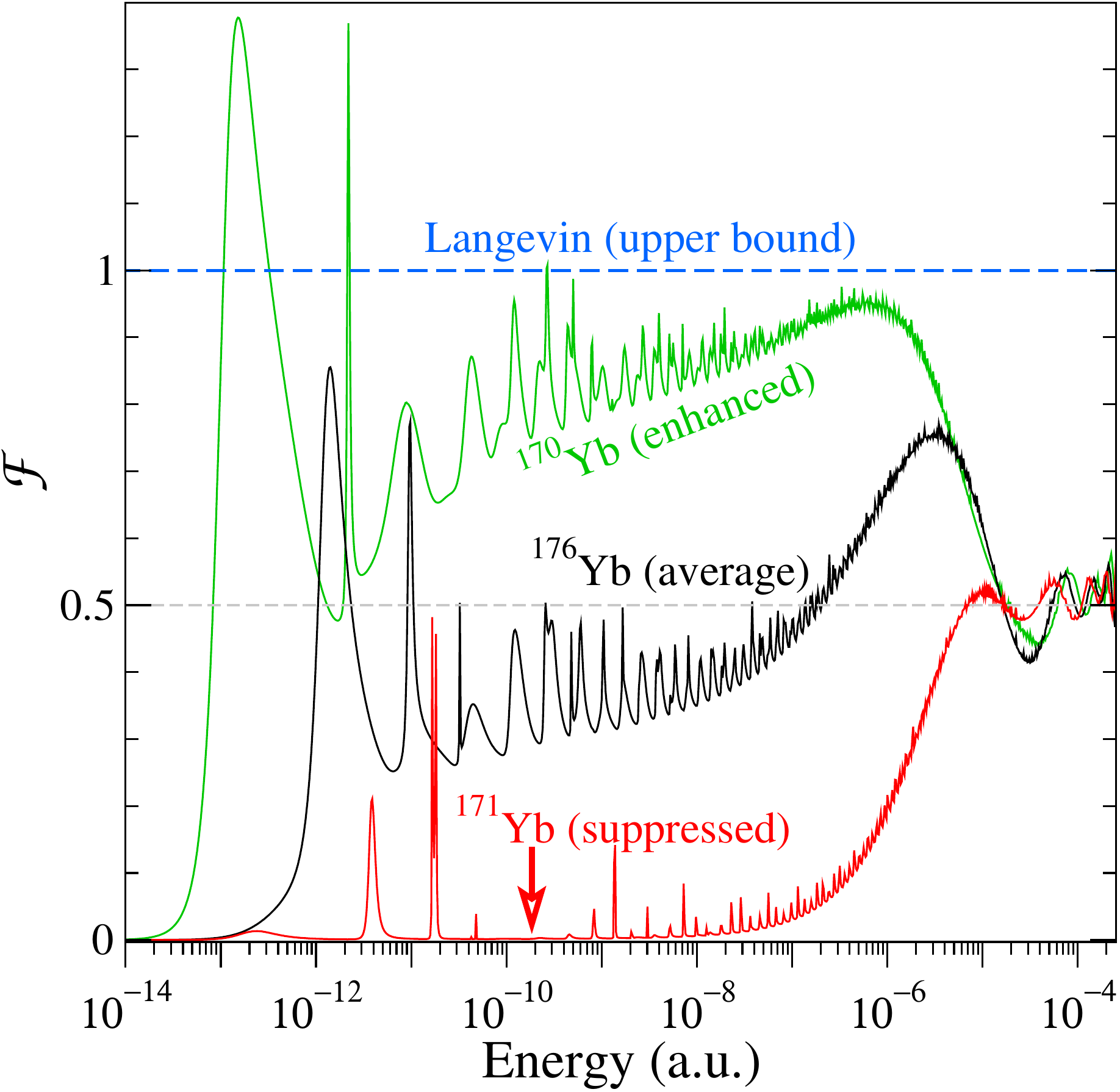}
  \caption{\label{fig:3-cases-semilog}
    Examples of enhanced, average, and suppressed cases.
  }
\end{figure}

\begin{figure}[t]
  \includegraphics[width=1.0\linewidth]{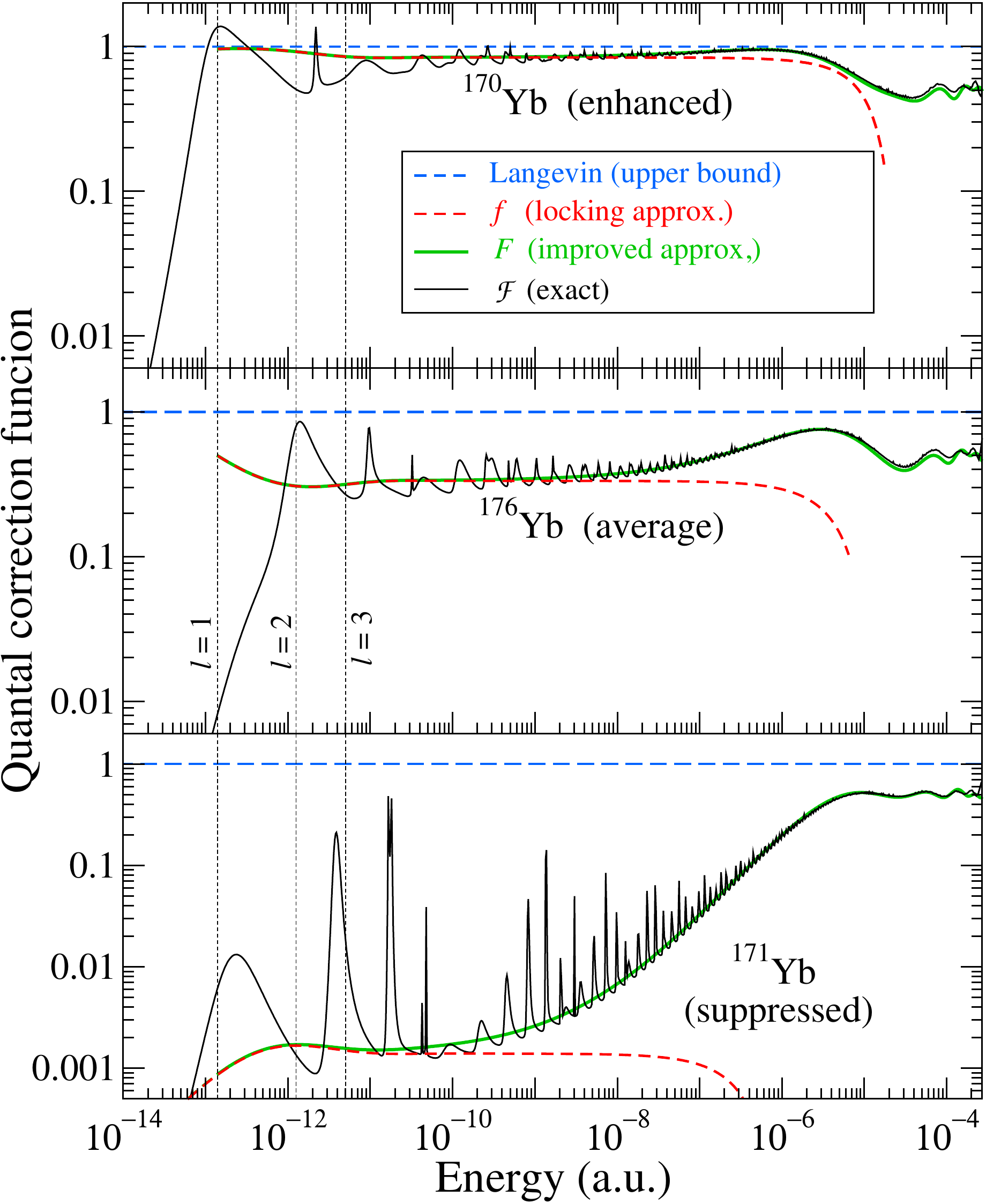}
\caption{\label{fig:F-3-panels}
  Quantal correction functions for the three different cases 
  (``enhanced'', ``average'', and ``suppressed'') depicting the energy where each partial
  wave $\ell$ starts contributing. The log-log scales help comparing $\mathcal{F}(E)$
  and the approximations $f(E)$ and $F(E)$ (the vertical scale is wider for the suppressed case).
  }
\end{figure}

%For a better understanding of
We can conveniently summarize the different behaviors according to the three cases
discussed above; namely, small values, say $|\Delta\delta_0|\lesssim\frac\pi6$,
correspond to suppression, while mid-range values
$\left(\frac\pi6\lesssim|\Delta\delta_0|\lesssim\frac\pi3\right)$ represent average
cases, and finally larger values
$\big(\frac\pi3\lesssim|\Delta\delta_0|\lesssim\frac\pi2\big)$ are enhanced cases.
Accordingly, this partitioning yields the following intervals for the
values of the quantal correction function: $F\lesssim\frac1 4$ for
suppression, $\frac14\lesssim F\lesssim\frac34$ for average, and
$\frac34 \lesssim F\lesssim 1$ for enhanced cases.  Although the intervals boundaries 
are somewhat arbitray, such a 
partitioning helps to distinguishing between the various behaviors. Fig.~\ref{fig:3-cases-semilog}
depicts examples of the three cases, showing the exact quantal function ${\cal F}$, while
Fig.~\ref{fig:F-3-panels} illustrate the same cases together with the approximate quantal functions
$f$ and $F$, pointing to the range of validity of the locking regime. It shows that the locking 
regime where $f$ is adequate covers a large range of energies for the enhanced case 
(up to $E\sim 10^{-7}$ a.u.), which decreases to roughly $10^{-8}$ a.u. for the average case, 
and shrinks to about $10^{-10}$ a.u. for the suppressed case.

\begin{figure}[h]
  \includegraphics[width=1.0\linewidth]{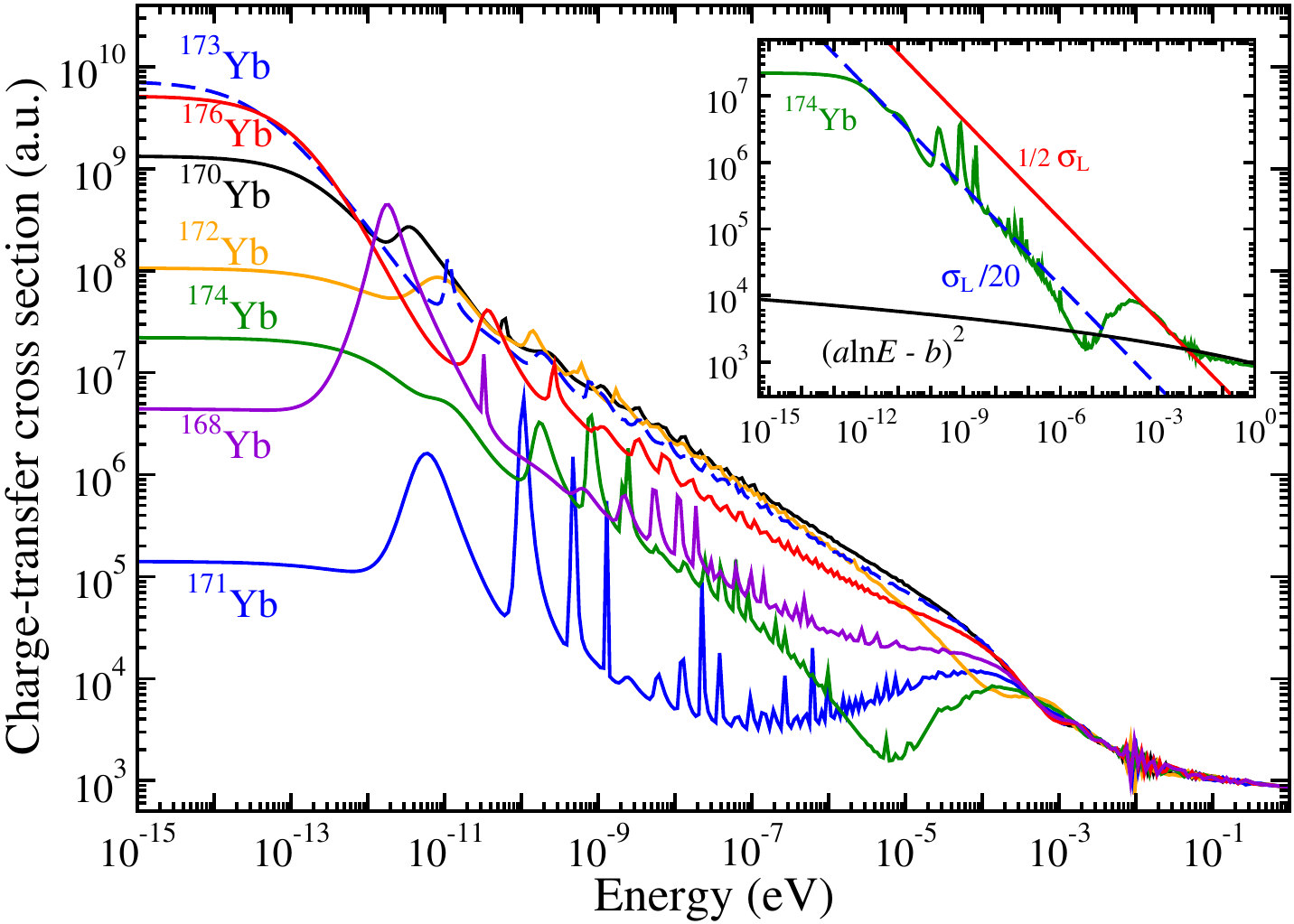}
\caption{\label{fig:Fig_4_new-final}
              Charge-transfer cross sections of resonance ion-atom collisions for stable Yb isotopes. The inset
              reproduces the results for $^{174}$Yb to illustrate the various energy regimes of $\sigma_{\rm exc}$,
              with $\sigma_{\rm exc}\approx \frac{1}{20}\sigma_L$ in the locking regime. It also shows the high-energy
              behavior $(a \ln E -b)^2$ not discussed here (see \cite{Yb-PRA-2009} for details; modified from 
              \cite{Yb-PRA-2009} where Langevin referred to $\overline{\sigma}_L$: see text).
  }
\end{figure}

%\noindent
We emphasize that although the suppression and enhancement phenomena may
seem anomalous, they are common occurrences rather than rare accidents. This was
already apparent in our previous work~\cite{PRL-2018-Simbotin-Cote} on Yb with a 
multitude of isotopes displaying an
abundance of such ``anomalies''. 
Finally, we point out that the different cases linked to the variation of $\Delta\delta_0$ also explains
results from prior studies in which departures from the Langevin cross section and large variations
between isotopes were obtained, such as the original work on Na with Na$^+$ \cite{Cote-Dalgarno-2000} with
$\sigma_{\rm exc}\sim \frac{1}{4}\sigma_L$, the large difference between isotopes in Li with Li$^+$ \cite{Peng-Li,PRA-Li-Li+2021}, or Yb~\cite{Yb-PRA-2009} (see also Fig.~\ref{fig:Fig_4_new-final}).
%\newpage

\subsection{
           The unlocking regime}
\label{sec:unlock}

As we discussed in previous sections, the locking of the phase shifts manifests itself
by $\Delta \eta_\ell \approx \mbox{constant}$, and their unlocking takes place when this
behavior is not followed. This occurs as $\ell$ grows, with the
$\lambda \cdot \Delta A_0=\ell (\ell +1) \cdot \Delta A_0$ contribution to $\Delta \eta_\ell$ 
becoming sizable (i.e., the condition $\lambda \ll |\frac{\Delta\delta_0}{\Delta A_0}|$ is not satisfied), 
as illustrated in Fig.~\ref{fig:Delta_eta_ell}  
for a given energy $E=10^{-5}$ a.u. (see Sec.~\ref{sec:lock}). Fig.~\ref{fig:eta-E} shows $\Delta \eta_\ell$
as a function of $E$ and its departure from the locking regime around $E\sim 6\times 10^{-6}$ a.u., roughly
the same range as $\Delta\delta_0$ starts to increase (see Fig.~\ref{fig:Deta-lock}).

The phase unlocking regime is reached when the locking approximation $f(E)$ 
in Eq.~(\ref{eq:f-lock}) starts failing as $E$ increases. This is apparent in the cross 
sections (Fig.~\ref{fig:sigma-regimes}) and
in the quantal correction function (Fig.~\ref{fig:F-3-panels}).
It is easily explained using the expression for $F(E)$ in Eq.~(\ref{eq:F}); when the value
of $\Lambda\cdot\Delta A_0$ becomes large enough as $E$ increases, it overcomes the
contribution of $\Delta \delta_0$ so that $\cos(2\Delta\delta_0 - \Lambda\cdot\Delta A_0)\napprox \cos(2\Delta\delta_0)$, rendering the approximation $F\approx f$ invalid.

We note that $\Delta A_0 \sim \mbox{constant}$ up to $10^{-5}$ a.u. (see Fig.~\ref{fig:DA0}) and
starts varying only when $E$ reaches roughly 1\% of $V_{\rm depth}$ of the shallowest potential 
curve $V_i$ (for Yb systems, $V_{\rm depth}\approx 6\times 10^{-4}$ a.u.; see Sec.~\ref{sec:theory-example}).
This emerges from the cancellation of the long-range contributions to $A_0^i$ from both channels (see Eq.~(\ref{eq:DA_ell-exact}) for $\Delta A_\ell$), leaving only the short-range contributions which vary 
little unless $E$ reaches values close to $V_{\rm depth}$. With both $\Delta A_0$ and $\Delta \delta_0$
varying little up to $E\sim 10^{-6}$ a.u., the onset of the unlocking regime arises from the growth of 
$\Lambda(E)\propto  E^{\frac{n-2}n}$ (see Eq.~(\ref{eq:LLn}):  $\Lambda(E)\propto \sqrt{E}$ for $n=4$).
The exact energy where unlocking takes place then depends entirely on the value of $\Delta \delta_0$; 
the split of $F$ from $f$ moves from higher energies for enhanced cases (e.g., $E\sim 10^{-7}$ a.u. for
$^{170}$Yb with $|\Delta \delta_0| \approx \frac{3\pi}{8}$) to lower energies 
for average cases (e.g., $E\sim 10^{-8}$ for $^{176}$Yb with $|\Delta \delta_0| \approx \frac{\pi}{4}$) 
and to even lower energies for suppressed cases (e.g., $E\sim 10^{-10}$ a.u. for $^{171}$Yb with 
$|\Delta \delta_0| \approx .01$).

Finally, as energies keep increasing, $F$ changes ``suddenly" before reaching an oscillatory behavior 
centered around $F \approx \frac{1}{2}$. 
%
%\marginpar{\color{black} \tiny There isn't an inset in Fig.5 (not needed). So, if we show a linear plot for F(E)
% then we show it as a separate Figure, not an inset in Fig.5 or 13, etc... }
%
Fig.~\ref{fig:F-exact-semilog} shows this
behavior for all Yb isotopes studied here. $F$ starts changing ``rapidly" around $10^{-7}-10^{-6}$ a.u. 
(depending on the isotope) and oscillates with an amplitude diminishing as 
$\frac1\Lambda\sim\sqrt{\frac{E_*}E}$ (see Eq.~(\ref{eq:LLn}) for $n=4$) 
centered about $\frac{1}{2}$. Using Eq.~(\ref{eq:F}) for $F$, we can rewrite 
$\sigma_{\rm exc}=\sigma_0+F\cdot \sigma_L$ in Eq.~(\ref{eq:sigma=sigma_0+Fsigma_L}) 
and explicitly express this oscillatory behavior as
\begin{equation}
  \sigma_\text{exc}  = \sigma_0+\frac{\sigma_L}2 - \sigma_\text{osc} ,
\end{equation}
with $\sigma_{\rm osc} \equiv \sigma_L \frac{\sin(\Lambda \cdot\Delta A_0)}{2\Lambda \cdot\Delta A_0}
 \cos(2\Delta\delta_0 - \Lambda \cdot\Delta A_0)$.
This result shows that the $s$-wave phase-shift difference
$\Delta\delta_0$ still affects $\sigma_{\rm exc}$ high above the Wigner
regime, impacting the exact oscillation phase.  However, as the energy increases, the dominant role in
modulating the energy dependence of $\sigma_\text{exc}$ relative to
its background value of $\frac1{2}\sigma_L$ is played by the
term $\Lambda \cdot \Delta A_0$.

Fig.~\ref{fig:Fig_4_new-final} is adapted from~\cite{Yb-PRA-2009} and summarizes the results, 
including the higher energy behavior where the current treatment is not adequate. 
This figure shows the ``exchange” regime"  $\sigma_{\rm exc} = (a\ln E-b)^2$ which would eventually 
reach the $E^{-6}$ of the Brinkman-Kramers cross section~\cite{Yb-PRA-2009}. We note that some 
authors define the Langevin cross section by including the $\langle \sin^2 \Delta \eta_\ell \rangle \approx \frac{1}{2}$
into their definition, {\it i.e.} $\overline{\sigma}_L = \frac{1}{2}\sigma_L$. In that case, one would define 
$\overline{\cal F} = 2 {\cal F}\in [0,2]$, with $\overline{\cal F} \rightarrow 1$ at higher energies and reaching a maximum value of 2. 

\section{Conclusion}
\label{sec:conclusion}

In the work, we explored how quantal cross sections can be related to classical cross sections.
We focused our attention on resonant-exchange processes, and formulated a quantal correction 
function ${\cal F}$ applied to the Langevin cross section $\sigma_L$. Using the properties of
phase shifts for different partial waves, we derived an approximate analytical expression $F$ for 
the quantal correction function that depends only on the $s$-wave phase shifts difference 
$\Delta \delta_0$ and the barrier height and difference in interaction potentials parametrized by 
$\Lambda$ and $\Delta A_0$, respectively. 

We applied the formulation to the case of resonant 
charge-transfer for Yb, and found three regimes where the cross section follows different behaviors, 
labeled Wigner, locking, and unlocking regimes. As the scattering energy increases, the cross section
migrates from the $s$-wave Wigner regime to the locking regime, where the phase shifts difference 
$\Delta \eta_\ell \approx \Delta \delta_0$ (modulo $\pi$) and $F\approx f = \sin^2\Delta \delta_0$, and 
to the unlocking regime where $\Delta \eta_\ell$ ceases to be roughly constant and start
varying with $\ell$ and $E$ due to the increasing contribution of $\Lambda\cdot\Delta A_0$.
Within the locking regime, we identified three distinct cases, dubbed suppressed, averaged, and enhanced,
corresponding to small (say $|\Delta\delta_0|\lesssim\frac\pi6$), mid-range $\left(\frac\pi6\lesssim|\Delta\delta_0|\lesssim\frac\pi3\right)$, and larger $\big(\frac\pi3\lesssim|\Delta\delta_0|\lesssim\frac\pi2\big)$
values of $|\Delta\delta_0|$,; this partitioning yields $F\lesssim\frac1 4$, $\frac14\lesssim F\lesssim\frac34$, 
and $\frac34 \lesssim F\lesssim 1$, respectively.

As noted previously \cite{PRL-2018-Simbotin-Cote}, the impact of $s$-wave scattering is apparent not only
within the Wigner regime (where it is expected to dominate), but at much higher energies where phase 
locking takes place. In addition to explaining large variations in resonant-exchange cross sections when
one would not expect $s$-waves to play any significant role with the contribution of large number of
partial waves, we furthermore find that $s$-waves dictate not only the range of the locking regime, but 
also its ``recovery" to reach the asymptotic value $F\approx \frac{1}{2}$. We found that it explains the unset 
of an oscillatory behavior $\sigma_{\rm osc}$ whose details still depend on $\Delta\delta_0$.

Finally, we stress that the results obtained here apply to any resonant-exchange process described
by the difference in phase shifts between two channels. It can be applied to various resonant processes
like charge-transfer, spin-flip, and excitation exchange, and for different interaction tails behaving as 
$r^{-n}$ covering the most common powers. In addition, the expression should be
applicable to quasi-resonant processes \cite{mott-massey}, like charge-transfer 
with mixed isotopes \cite{Peng-Li, Peng-Be}, reactions
involving different hyperfine asymptotes \cite{rui2017controlled},
isotope substitutions \cite{Bala-H2+D,PCCP-H2+D}, or atom exchange \cite{Ni-Science-2024,Byrd-2010},
as long as the scattering energy is larger than the energy gap between the asymptotes
of the relevant channels. 
The influence of $s$-wave scattering at higher temperatures also provide a diagnostic tool 
particularly relevant to system for which ultracold temperatures $T$ are not easily achievable, such as
ion-atom hybrid systems for which the nK regime remains a challenge, or in ultracold chemical
systems. Indeed, by measuring the cross section or rate for a resonant process at higher 
temperatures more easily accessible ({\it e.g.}, spin-flip, charge-transfer, or substitution reaction), 
one can gain information about the $s$-wave regime. Observed sizable suppression compared to
$\sigma_L$ implies that the $s$-wave phase shifts are close to each other and $\sigma_{\rm exc}$
will remain small at lower energies, while large values will remain large at low energies. This can guide
experiments in identifying the system to investigate; selecting systems with small values at higher $T$ 
to study system where exchange processes can be minimized if there are a nuisance ({\it e.g.}, 
in atom-ion interaction to maintain specific superpositions of internal states), or selecting systems
with large values if they are in fact required ({\it e.g.}, for cooling ions by charge-transfer with an ultracold 
buffer gas).

\color{black}

\begin{acknowledgments}
  This work was partially supported by the National Science Foundation
  Grant PHY-2513221 and OSI-2328774.
\end{acknowledgments}
   
%
%\color{violet}%
%
\bibliography{res-exch-F-QM}
\end{document}